\documentclass[11pt,a4paper]{article} 
\usepackage{jheppub,hyperref,mdwlist}
\usepackage{amsmath}
\usepackage{graphicx}
\usepackage{subfigure}
\usepackage{amssymb}
\usepackage{xcolor}
\usepackage{multirow}
\usepackage{cancel}
\usepackage{color}
\usepackage{listings}
\usepackage{xcolor}
\usepackage{float}
\usepackage{slashed}

\usepackage{amsmath}
\usepackage{wrapfig}
\usepackage{cleveref}
\usepackage{mathtools}

\newcommand{\be}{\begin{equation}}
\newcommand{\ee}{\end{equation}}
\newcommand{\beq}{\begin{equation}}
\newcommand{\eeq}{\end{equation}}
\newcommand{\bea}{\begin{eqnarray}}
\newcommand{\eea}{\end{eqnarray}}
\newcommand{\besp}{\begin{equation}\begin{split}}
\newcommand{\eesp}{\end{split}\end{equation}}

\newcommand{\Eq}[1]{Eq.~(\ref{#1})}

\newcommand{\Dfbd}{\mathord{\buildrel{\lower3pt\hbox{$\scriptscriptstyle\leftrightarrow$}}\over {D}_{\mu}}}

\def\mL{\mathcal{L}}

\def\mO{\mathcal{O}}

\def\0{\textbf{0}}
\def\1{\textbf{1}}
\def\2{\textbf{2}}
\def\3{\textbf{3}}
\def\4{\textbf{4}}
\def\5{\textbf{5}}
\def\6{\textbf{6}}
\def\7{\textbf{7}}
\def\8{\textbf{8}}
\def\9{\textbf{9}}

\def\d{\text{d}}

\begin{document}

\title{Pinning down the primordial black hole formation mechanism with gamma-rays and gravitational waves}

\author{Ke-Pan Xie}

\affiliation{School of Physics, Beihang University, Beijing 100191, China}

\emailAdd{kpxie@buaa.edu.cn}

\abstract{

Primordial black holes (PBHs) are predicted in many models via different formation mechanisms. Identifying the origin of PBHs is of the same importance as probing their existence. We propose to probe the asteroid-mass PBHs [$\mathcal{O}(10^{17})~{\rm g}\lesssim M\lesssim\mathcal{O}(10^{22})~{\rm g}$] with gamma-rays from Hawking radiation and the stochastic gravitational waves (GWs) from the early Universe. We consider four concrete formation mechanisms, including collapse from primordial curvature perturbations, first-order phase transitions, or cosmic strings, and derive the extended PBH mass functions of each mechanism for phenomenological study. The results demonstrate that by combining gamma-rays and GW signals we can probe PBHs up to $\mathcal{O}(10^{19})~{\rm g}$ and identify their physical origins.

}

\maketitle
\flushbottom

\section{Introduction}

Many models predict the formation of black holes in the early Universe, soon after the Big Bang. Those {\it primordial} black holes (PBHs), in contrast to ``astrophysical" black holes that are formed from stellar collapses, exist long before the formation of galaxies and stars~\cite{zel1967hypothesis,hawking1971gravitationally}. Therefore, the mass of PBHs is not necessarily related to the stellar mass and can be in a vast range, depending on the formation mechanism. Different mass ranges receive experimental probes from different astrophysical or cosmological observations. Due to the Hawking radiation~\cite{Hawking:1974rv}, PBHs with mass $M\lesssim5\times10^{14}$ g would have evaporated before today and could leave their imprints in the Big Bang Nucleosynthesis (BBN)~\cite{Carr:2009jm}, Cosmic Microwave Background (CMB)~\cite{Acharya:2020jbv, Chluba:2020oip}, extragalactic gamma-rays~\cite{Carr:2009jm}, or gravitational waves~\cite{Papanikolaou:2020qtd,Papanikolaou:2022chm,Papanikolaou:2021uhe,Papanikolaou:2022hkg}. Heavier PBHs can survive till today and be probed by gamma-rays (for $M\lesssim10^{19}$ g), gravitational microlensing (for $M\gtrsim10^{21}$ g), accretion (for $10^{34}~{\rm g}\lesssim M\lesssim 10^{41}~{\rm g}$), etc. See Refs.~\cite{Carr:2020gox,Carr:2020xqk,Green:2020jor} for reviews on current constraints on PBHs.

Depending on their masses, PBHs have very rich cosmological implications. Light PBHs can Hawking evaporate into dark matter~(DM)~\cite{Bell:1998jk, Khlopov:2004tn, Allahverdi:2017sks, Lennon:2017tqq, Gondolo:2020uqv, Masina:2020xhk, Bernal:2021yyb, Cheek:2021odj, Cheek:2021cfe, Sandick:2021gew, Cheek:2022mmy}, dark radiation~\cite{ Hooper:2019gtx, Masina:2021zpu,Arbey:2021ysg,Cheek:2022dbx} or generate  the baryon asymmetry of the Universe~\cite{Toussaint:1978br, Turner:1979bt, Grillo:1980rt, Baumann:2007yr, Fujita:2014hha, Hook:2014mla,Hamada:2016jnq, Morrison:2018xla, Bernal:2022pue,Hooper:2020otu,Perez-Gonzalez:2020vnz,Datta:2020bht,JyotiDas:2021shi,Gehrman:2022imk}. PBHs with tens to hundreds of the solar mass ($M_\odot$) might explain the merger signals observed by the LIGO/Virgo collaborations~\cite{LIGOScientific:2016aoc,LIGOScientific:2016sjg,LIGOScientific:2017bnn,Clesse:2016vqa,Bird:2016dcv,Sasaki:2016jop}. Even heavier PBHs, with $M\gtrsim\mO(10^3-10^5)M_\odot$, could seed the superheavy black holes~\cite{Bean:2002kx,Khlopov:2004sc,Duechting:2004dk,Kawasaki:2012kn,Clesse:2015wea, DeLuca:2022bjs}. Especially, PBHs with $\mO(10^{17})~{\rm g}\lesssim M\lesssim\mO(10^{22})~{\rm g}$, known as the ``{\it asteroid-mass range}'', can still explain all the DM abundance~\cite{Carr:2020xqk}. It is well known that asteroid-mass PBHs can be probed by Hawking radiation.\footnote{There are also other approaches to probe this mass range, such as gamma-ray burst lensing~\cite{Jung:2019fcs}.} Indeed, the existing astronomical observations on gamma-rays~\cite{Laha:2020ivk,Coogan:2020tuf,Laha:2019ssq,DeRocco:2019fjq}, $e^\pm$ and neutrinos~\cite{Boudaud:2018hqb,Dasgupta:2019cae} have put stringent bounds on $f_{\rm pbh}$, i.e. the fraction of DM contributed by PBHs. For a monochromatic PBH mass function, the upper bound on $f_{\rm pbh}$ varies approximately from $10^{-8}$ to 1 for $M$ ranging from $10^{15}$ g to $10^{17}$ g, with an $M^4$ scaling~\cite{Carr:2020gox,Carr:2020xqk,Green:2020jor}. Future gamma-ray detectors are able to probe PBH DM candidate up to $\mO(10^{19})~{\rm g}$~\cite{Coogan:2020tuf,Ray:2021mxu,Ghosh:2021gfa}.

In this article, we propose to probe asteroid-mass PBHs with near-future gamma-ray detectors and gravitational wave (GW) detectors. Instead of phenomenologically assuming a monochromatic PBH mass function, we consider concrete physical mechanisms of PBH formation, which yield extended mass functions and hence predict more realistic gamma-ray signal spectra. The four mechanisms under consideration are
\begin{enumerate}
\item Collapse of overdense regions originating from curvature perturbations generated during inflation~\cite{Carr:1974nx,Carr:1975qj};
\item Direct collapse of false vacuum remnants during a cosmic first-order phase transition (FOPT)~\cite{Baker:2021nyl,Baker:2021sno};
\item Subsequent collapse of non-topological solitons which are formed in a FOPT~\cite{Kawana:2021tde};
\item Collapse of cosmic strings~\cite{Hawking:1987bn}.
\end{enumerate}
Remarkably, all those mechanisms have stochastic GW companions. This provides us the opportunity to probe the origin of PBHs via multi-messenger astronomy. An analysis of correlating gamma-ray and GWs from the curvature perturbation PBH mechanism is performed in Ref.~\cite{Agashe:2022jgk}, showing the possibility of testing the above formation mechanism 1. In current article, we for the first time discuss the comparison and identification of signals from different PBH mechanisms via multi-messenger astronomy.

This work discusses the main features of gamma-ray and GW signals from concrete formation mechanisms, demonstrating that they are qualitatively distinguishable. For a quantitative study, we consider the MeV gamma-ray detectors e-ASTROGAM~\cite{e-ASTROGAM:2016bph} and AMEGO-X~\cite{Fleischhack:2021mhc}, which are planed for launch at the end of the 2020s; and a few GW detectors, which are already under operation or proposed to start data-taking in the 2030s, including the pulsar timing arrays (PTAs) NANOGrav~\cite{McLaughlin:2013ira,NANOGRAV:2018hou,Aggarwal:2018mgp,Brazier:2019mmu}, PPTA~\cite{Manchester:2012za,Shannon:2015ect}, EPTA~\cite{Kramer:2013kea,Lentati:2015qwp,Babak:2015lua}, IPTA~\cite{Hobbs:2009yy,Manchester:2013ndt,Verbiest:2016vem,Hazboun:2018wpv} and SKA~\cite{Carilli:2004nx,Janssen:2014dka,Weltman:2018zrl}, the space-based laser interferometers LISA~\cite{LISA:2017pwj}, TianQin~\cite{TianQin:2015yph,Hu:2017yoc,TianQin:2020hid}, Taiji~\cite{Hu:2017mde,Ruan:2018tsw}, BBO~\cite{Crowder:2005nr} and DECIGO~\cite{Kawamura:2011zz}, and the ground-based interferometers LIGO~\cite{LIGOScientific:2014qfs,LIGOScientific:2019vic}, CE~\cite{Reitze:2019iox} and ET~\cite{Punturo:2010zz,Hild:2010id,Sathyaprakash:2012jk}. Our research shows that, after combining gamma-ray and GW signals, we can probe the existence of PBHs with mass up to $\mO(10^{19})~{\rm g}$, and equally importantly identify their physical origin.

This article is organized as follows. We first introduce the four PBH mechanisms one by one in Section~\ref{sec:inflation} (curvature perturbations), Section~\ref{subsec:2105PBH} (direct collapse during a FOPT), Section~\ref{subsec:solitons} (collapse of non-topological solitons from a FOPT), and Section~\ref{sec:cs} (cosmic strings), and discuss the corresponding gamma-ray and GW signals, as well as their correlations. Then in Section~\ref{sec:summary} we summarize our results with a combined discussion of different mechanisms, pointing out how to distinguish them in future experiments. The conclusion is also given.

\section{PBHs from curvature perturbations}\label{sec:inflation}

Large scalar perturbation generated by inflationary theories can source PBH formation~\cite{Carr:1975qj,Ivanov:1994pa,Garcia-Bellido:1996mdl,Silk:1986vc,Kawasaki:1997ju,Yokoyama:1995ex,Choudhury:2013woa,Di:2017ndc,Pi:2017gih,Hertzberg:2017dkh, Ballesteros:2017fsr,Cai:2018dig,Dalianis:2018frf,Ozsoy:2018flq, Cicoli:2018asa, ShamsEsHaghi:2022azq,Choudhury:2023vuj}. While the Universe is confirmed to be nearly homogeneous at large scales by measurements of Lyman-$\alpha$ forest~\cite{Bird:2010mp}, CMB anisotropy~\cite{Planck:2018jri} and CMB distortion~\cite{Mather:1993ij, Fixsen:1996nj}, small scale fluctuations are still less constrained. For example, the matter power spectrum of curvature perturbations $P_\zeta$ could be enhanced during inflation when the inflaton field undergoes a temporary ultra slow-roll phase~\cite{Martin:2012pe, Kinney:2005vj, Germani:2017bcs, Dimopoulos:2017ged, Riotto:2023hoz,Kawai:2021edk,Kawai:2021bye,Kawai:2022emp,Wang:2022nml}. The fluctuations, after being produced, become super-horizon modes and stay frozen until they enter the causal horizon as overdense regions much after the end of the inflation epoch. After horizon reentry, PBHs are formed from the direct collapse of dense horizon patches whose distribution are determined by $P_\zeta(k)$, where $k$ is the comoving wave number. Gravitational collapse happens when density contrast $\delta$ in the horizon becomes larger than the threshold value $\delta_c$. We adopt the Press-Schechter (PS) formalism~\cite{Press:1973iz} for the calculation of PBH abundance with the assumption that density contrasts in the early Universe follow a Gaussian distribution
\bea
p(\delta)=\frac{1}{\sqrt{2\pi\sigma^2_0}} \, e^{-\frac{\delta^2}{2\sigma^2_0}}.
\eea
Here the mean value of $\delta$ is zero, because the Universe is nearly homogeneous and the overdense and underdense regions appear with equal probabilities. The variance is given by
\be\label{eq:sigma0}
\sigma_0^2(R)=\displaystyle{\int_0^{\infty}}\frac{{\rm d}k'}{k'}\frac{16}{81}(k'R)^4W^2(k',R)P_\zeta(k'),
\ee
where $R=1/(Ha)$ is the comoving Hubble radius, with $H$ the Hubble rate and $a$ the scale factor. A window function $W(k',R)$ is used to smooth the density contrast, which we take to be Gaussian in this study
\be
W(k',R)=\exp\left(-\frac{(k'R)^2}{4}\right).
\ee
Since a comoving Hubble radius $R$ corresponds to a reentry wave number $k\equiv R^{-1}$, $\sigma_0^2$ can also be treated as a function of $k$ via \Eq{eq:sigma0}.

Horizon patches with $\delta>\delta_c$ collapse to PBHs with mass~\cite{Choptuik:1992jv,Niemeyer:1997mt,Niemeyer:1999ak}
\be\label{eq:MR}
M=M_HK(\delta-\delta_c)^{\gamma_r},
\ee
where $M_H$ is the horizon mass as a function of $R$, and we use $\delta_c=0.55$, $K=10$ and $\gamma_r=0.36$~\cite{Musco:2020jjb,Escriva:2021aeh}. The energy density ratio of PBH to radiation, $\beta_{\rm pbh}$, is
\be
\frac{\partial^2\beta_{\rm pbh}}{\partial\delta\partial\log R}=\frac{2M}{M_H}p(\delta),\quad \delta\geqslant\delta_c,
\ee
at the moment of PBH formation. To get the PBH distribution today, we take into account the cosmic expansion, reheating from the decoupling of Standard Model (SM) particle species and the PBH mass loss via Hawking evaporation,
\be\label{eq:massfunctioninflation}
\frac{\d f_{\rm pbh}}{\d M}=\frac{\Omega_mh^2}{\Omega_{\rm DM}h^2}\left(\frac{M}{M'}\right)^3\frac{\d\beta_{\rm pbh}^{\rm eq}}{\d M}\Big|_{M\to M'},
\ee
where $M'=(M^3+3f_0t_0M_{\rm Pl}^4)^{1/3}$, and
\be
\frac{\d\beta_{\rm pbh}^{\rm eq}}{\d M}=\int_0^{\rm R_{\rm eq}}\frac{\d R}{R}\frac{R_{\rm eq}}{R}\left(2\frac{g_*(T_i)}{g_*(T_{\rm eq})}\frac{g_{*s}^{4/3}(T_{\rm eq})}{g_{*s}^{4/3}(T_i)}\right)^{1/2}\frac{\partial^2\beta_{\rm pbh}}{\partial\delta\partial\log R}\frac{\d\delta}{\d M},
\ee
is the ratio of PBH energy density to radiation at matter-radiation equality~\cite{Agashe:2022jgk}. $g_*$ and $g_{*s}$ are the numbers of relativistic degrees of freedom for energy and entropy, respectively, and the variables labeled with subscript ``eq'' and ``$i$'' are given at matter-radiation equality and PBH formation time, respectively. The abundances of matter and cold DM are $\Omega_mh^2\approx0.142$ and $\Omega_{\rm DM}h^2\approx0.120$, respectively~\cite{ParticleDataGroup:2020ssz}. The PBH evaporation parameter is $f_0=1.895\times10^{-3}$~\cite{Hooper:2019gtx} and the age of the present Universe is $t_0=4.6\times10^{17}~{\rm s}$.

Given a curvature perturbation $P_\zeta(k)$, one can derive $\d f_{\rm pbh}/\d M$ via the above procedure. If $P_\zeta(k)$ is enhanced at some specific $k$-mode $k_p$, when the enhanced mode reenters the horizon, the variance $\sigma_0(R)$ is also enhanced and hence large density contrast could exist. This generates a peak in the PBH mass function. In this article, we parametrize the curvature perturbation using a log-normal function
\be\label{eq:deltaPzeta}
P_{\zeta}(k)=\frac{A}{\sqrt{2\pi\sigma^2}} \, \exp\left(\frac{\log^2 \left(k/k_p\right)}{2\sigma^2}\right).
\ee
By this setup, given a set of $(A,k_p,\sigma)$, one can derive the mass function $\d f_{\rm pbh}/\d M$. The non-linear relation between the density contrast and curvature perturbations requires $\sim2$ larger $P_{\zeta}$, and this extra factor is included in our calculation~\cite{Young:2019yug}. Once $\d f_{\rm pbh}/\d M$ is available, the gamma-ray spectrum can be evaluated with the method described in Appendix~\ref{app:gamma-ray}.

\begin{figure}
\centering
\includegraphics[scale=0.39]{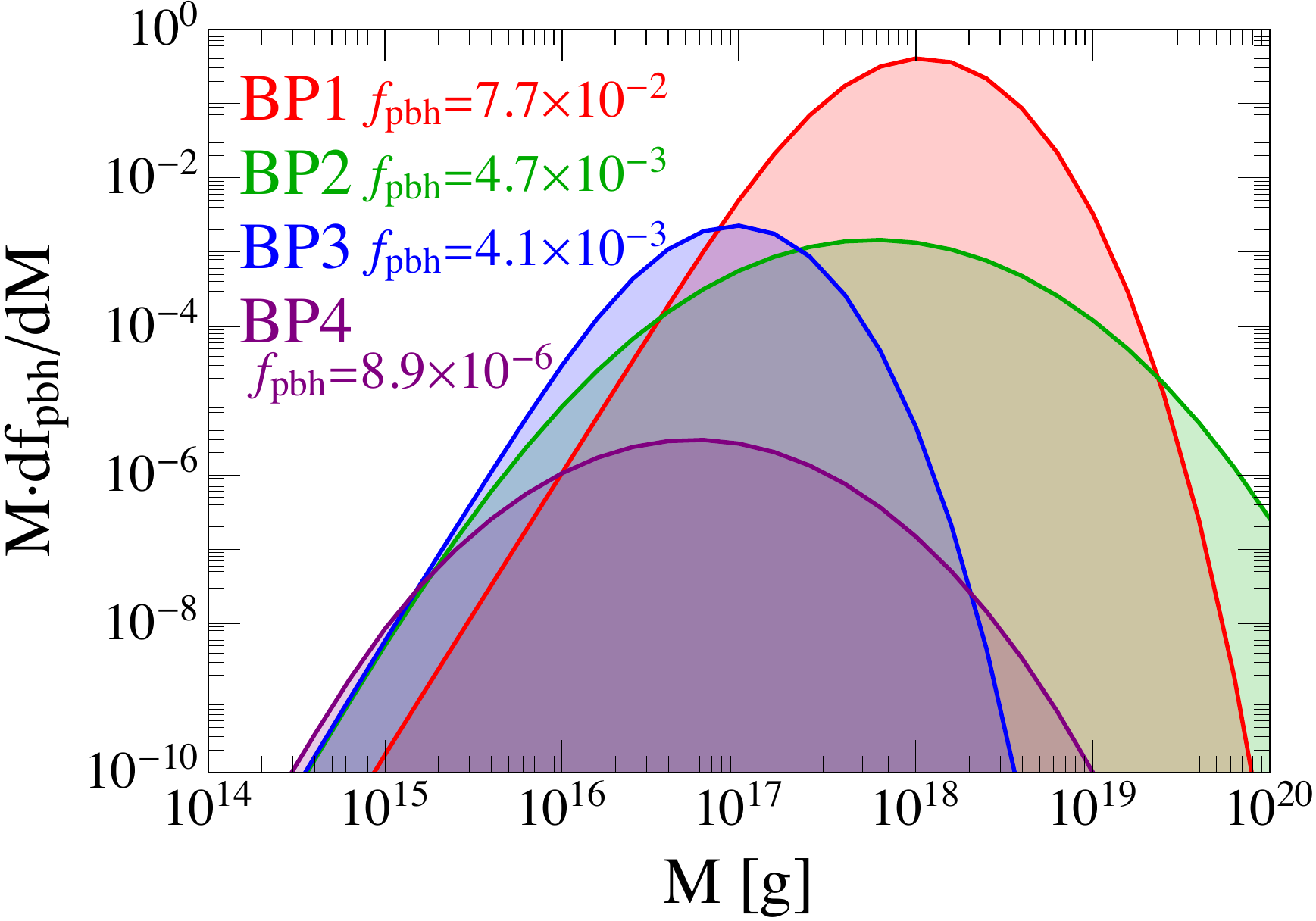}\qquad
\includegraphics[scale=0.4]{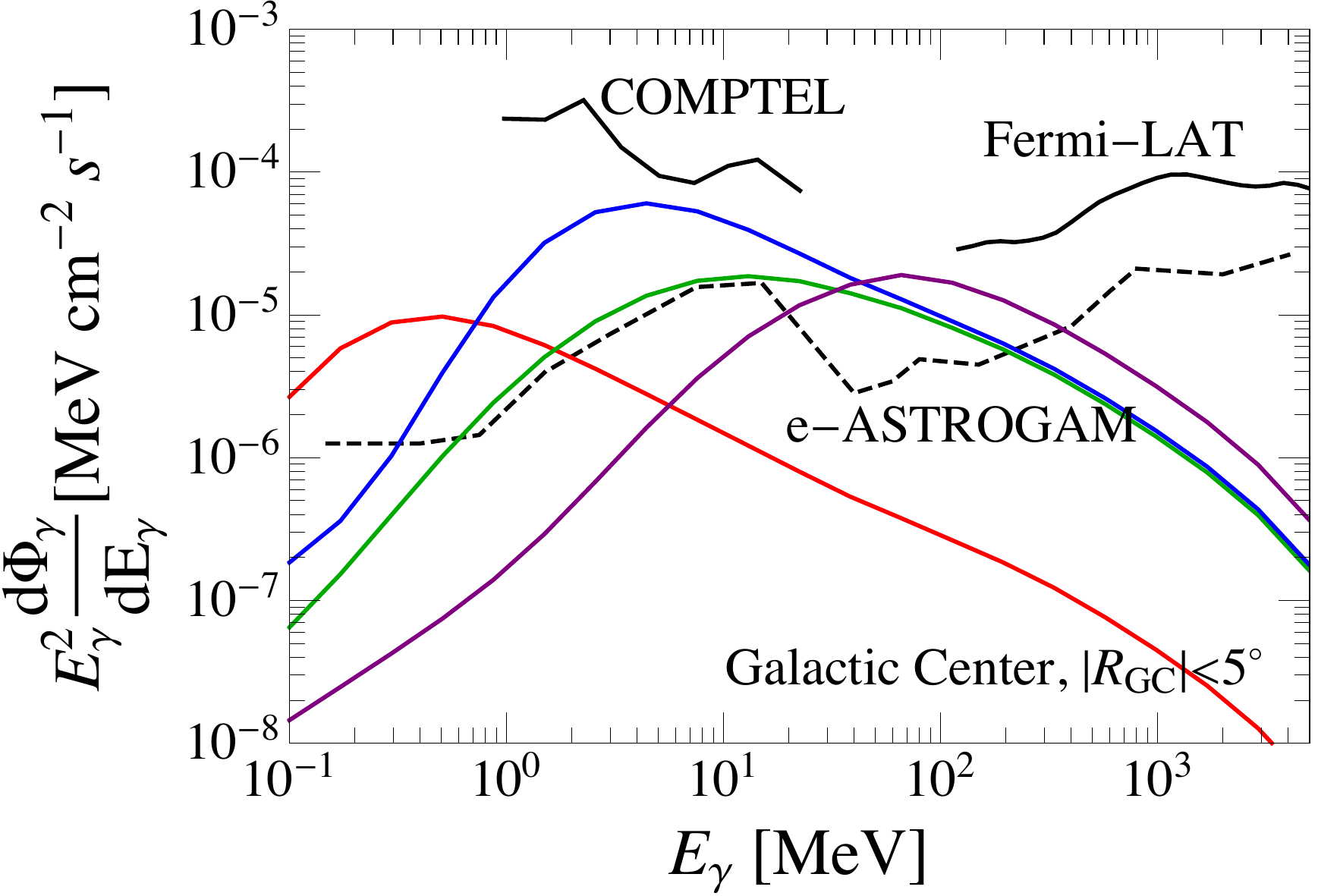}
\caption{The current mass distributions ({\bf left}) and corresponding gamma-ray spectra ({\bf right}) of the PBHs from curvature perturbations. The description of the BPs can be found in \Eq{BPs_inflation}.}
\label{fig:inflation}
\end{figure}

The input parameters $(A,k_p,\sigma)$ affect the PBH mass function in a very clear way: $A$ controls $f_{\rm pbh}$, $k_p$ determines the peak position of $M$, while $\sigma$ dominates the width of the mass peak, as shown in the parameter scan of Ref.~\cite{Agashe:2022jgk}. For four benchmark points (BPs)
\be\label{BPs_inflation}
\{A,~k_p,~\sigma\}=\begin{cases}~\{10^{-1.26},~3\times10^{14}~{\rm Mpc}^{-1},~2\},&{\rm BP1};\\
~\{10^{-1.03},~3\times10^{14}~{\rm Mpc}^{-1},~4\},&{\rm BP2};\\
~\{10^{-1.33},~10^{15}~{\rm Mpc}^{-1},~2\},&{\rm BP3};\\
~\{10^{-1.10},~10^{15}~{\rm Mpc}^{-1},~4\},&{\rm BP4},\\
\end{cases}
\ee
we plot the PBH mass functions in the left panel of Fig.~\ref{fig:inflation}. The BPs are selected based on the consideration that they should produce asteroid-mass PBHs, and should not be ruled out by current gamma-ray observations, and have $f_{\rm pbh}\leqslant1$. This leads to BPs with $k_p\sim10^{15}~{\rm Mpc}^{-1}$ with various $A$ and $\sigma$ values. We can see the mass peaks match the estimation in \Eq{eq:MR}, and a larger $\sigma$ of the power spectrum yields a broader PBH mass distribution.  The corresponding gamma-ray spectra are plotted in the right panel of 
Fig.~\ref{fig:inflation}, where the signals are from the galactic center with an angle extent $|R_{\rm GC}|\leqslant5^\circ$. In the same figure, we also add the current constraints from gamma-ray observations Fermi-LAT~\cite{Fermi-LAT:2017opo},  COMPTEL~\cite{1998PhDT.........3K} and the projections from e-ASTROGAM~\cite{e-ASTROGAM:2016bph}, all rescaled to the region of interest (ROI) of our study. In particular, we take the COMPTEL constraint from~\cite{Essig:2013goa}. The projected reach of e-ASTROGAM is derived by rescaling the expected background numbers provided by Ref.~\cite{e-ASTROGAM:2016bph} to our ROI and requiring a $2\sigma$ deviation contributed by the signals in each bin. We expect the AMEGO-X~\cite{Fleischhack:2021mhc} detector has a similar sensitivity.

Curvature perturbations can source stochastic GWs via tensor mode produced at the second-order. The effect of induced GWs from scalar perturbations is studied in Refs.~\cite{1967PThPh..37..831T, Mollerach:2003nq, Ananda:2006af, Baumann:2007zm, Acquaviva:2002ud, Yuan:2021qgz, Domenech:2021ztg, Pi:2020otn, Kohri:2018awv, Espinosa:2018eve, Braglia:2020eai, Inomata:2019ivs, Inomata:2016rbd, Inomata:2018epa,Kozaczuk:2021wcl, Agashe:2022jgk}. Here we follow Refs.~\cite{Kozaczuk:2021wcl,Agashe:2022jgk} for the calculation in our analysis. The GW spectrum is defined as the GW energy density fraction as a function of the frequency,
\be
\Omega_{\rm GW}(f)=0.83  \left(\frac{g_{*s}(T_c)}{10.75}\right)^{-\frac{1}{3}}  \Omega_{r,0} \Omega_{\rm GW}(\eta_c,k),
\ee
with $\Omega_{r,0} = 8.5\times10^{-5}$ being the current radiation abundance, respectively; and $\eta_c$ and $T_c$ are the reentry conformal time and temperature determined by $k$, respectively. The $k$-mode is related to GW frequency via
\be\label{fk}
f_{\rm GW}=1.546~{\rm Hz}\times\left(\frac{k}{10^{15}~{\rm Mpc}^{-1}}\right),
\ee
which indicates the peak frequency of GWs is determined by the horizon reentry of enhanced $k$-mode of $P_{\zeta}(k)$.

The full expression of GW energy fraction at a given $k$ at conformal time $\eta$ is
\be
\Omega_{\rm GW}(\eta,k)=\frac{1}{24}\left(\frac{k}{a(\eta)H(\eta)}\right)^2P_h(\eta,k),
\ee
with the power spectrum of the GW being
\be
P_h(\eta,k)\simeq2\int^{\infty}_{0}\d t\int^{1}_{-1}\d s\left(\frac{t(t+2)(s^2-1)}{(t+s+1)(t-s+1)}\right)^2
I^2(s,t,k\eta)P_\zeta(u k)P_\zeta(v k).
\ee
The $I^2$ function can be calculated when the GWs are deeply in the sub-horizon limit. In that limit, the growth of GW amplitude is terminated by the decay of the enhanced $k$-mode once it is much smaller than the horizon size. In this limit, we can take
\be\begin{split}
I^2(s,t,k\eta)=&~\frac{288(s^2+t(t+2)-5)^2}{k^2\eta^2(t+s+1)^6(t-s+1)^6}\left[\frac{\pi^2}{4}(s^2+t(t+2)-5)^2\Theta(t-(\sqrt{3}-1))\right.\\
&~\left.+\left((t+s+1)(s-t-1)+\frac{s^2+t(t+2)-5}{2}\log\left|\frac{t(t+2)-2}{3-s^2}\right|\right)^2\right],
\end{split}\ee
where $\Theta$ is the Heaviside function.

\begin{figure}
\centering
\includegraphics[scale=0.4]{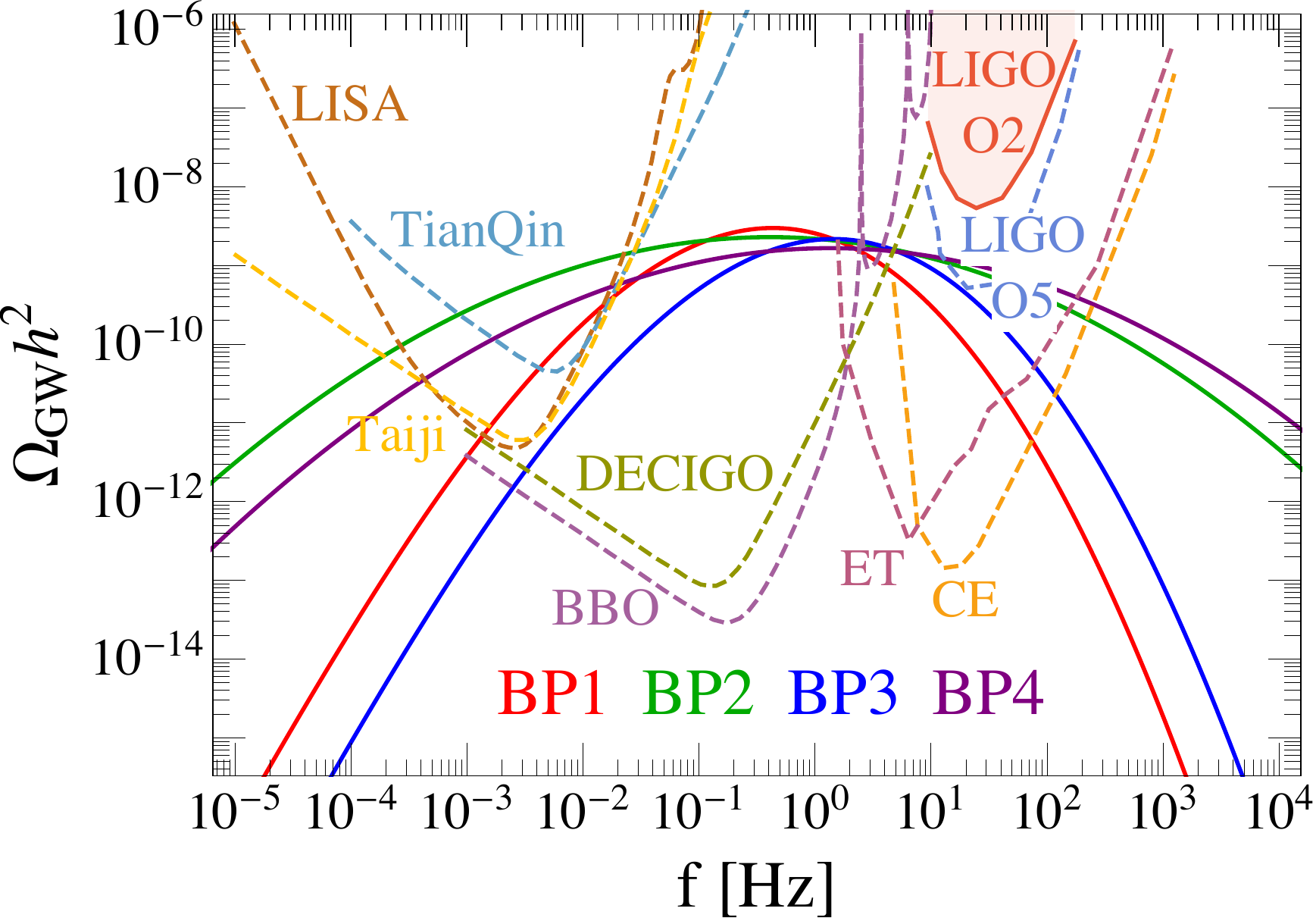}
\caption{The GW spectra of the BPs in \Eq{BPs_inflation}, which are correlated signals from PBHs induced by curvature perturbation.}
\label{fig:gw_inflation}
\end{figure}

Since the PBH mass is proportional to $k^{-2}$, the curvature perturbation responsible for the generation of heavy~(light) PBHs lies in the lower~(higher) frequency regions. For PBH mass in the asteroid-mass window, the target GW signals have $\sim\text{Hz}$ frequency, as implied by \Eq{eq:MR} and \Eq{fk}. The abundance of GWs $\Omega_{\rm GW}$ is quadratic in the amplitude of the power spectrum $P_\zeta(k)$. As the amplitude is required to be larger than $\sim 10^{-2}$ for sufficient PBH formation, the GWs are estimated to make up $10^{-4}$ of the energy density of the Universe at the horizon-reentry time, and then redshift to $10^{-9}$ in the current Universe. This is within the sensitive region of a few ground-based and proposed space-based GW detectors, as illustrated in Fig.~\ref{fig:gw_inflation}. The signals are best probed by the future BBO~\cite{Crowder:2005nr} and DECIGO~\cite{Kawamura:2011zz} detectors, but the near-future LISA~\cite{LISA:2017pwj}, TianQin~\cite{TianQin:2015yph,Hu:2017yoc,TianQin:2020hid}, Taiji~\cite{Hu:2017mde,Ruan:2018tsw}, CE~\cite{Reitze:2019iox} and ET~\cite{Punturo:2010zz,Hild:2010id,Sathyaprakash:2012jk}, or even the operating LIGO~\cite{LIGOScientific:2014qfs,LIGOScientific:2019vic}, also have considerable sensitivities to probe them. Combining the gamma-ray and GW signals, we can efficiently probe the curvature perturbation-induced PBH scenario, and this has been pointed out by Ref.~\cite{Agashe:2022jgk}.

\section{PBHs from a FOPT}\label{sec:FOPT}

\subsection{The general mechanism}

\begin{figure}
\centering
\includegraphics[scale=0.3]{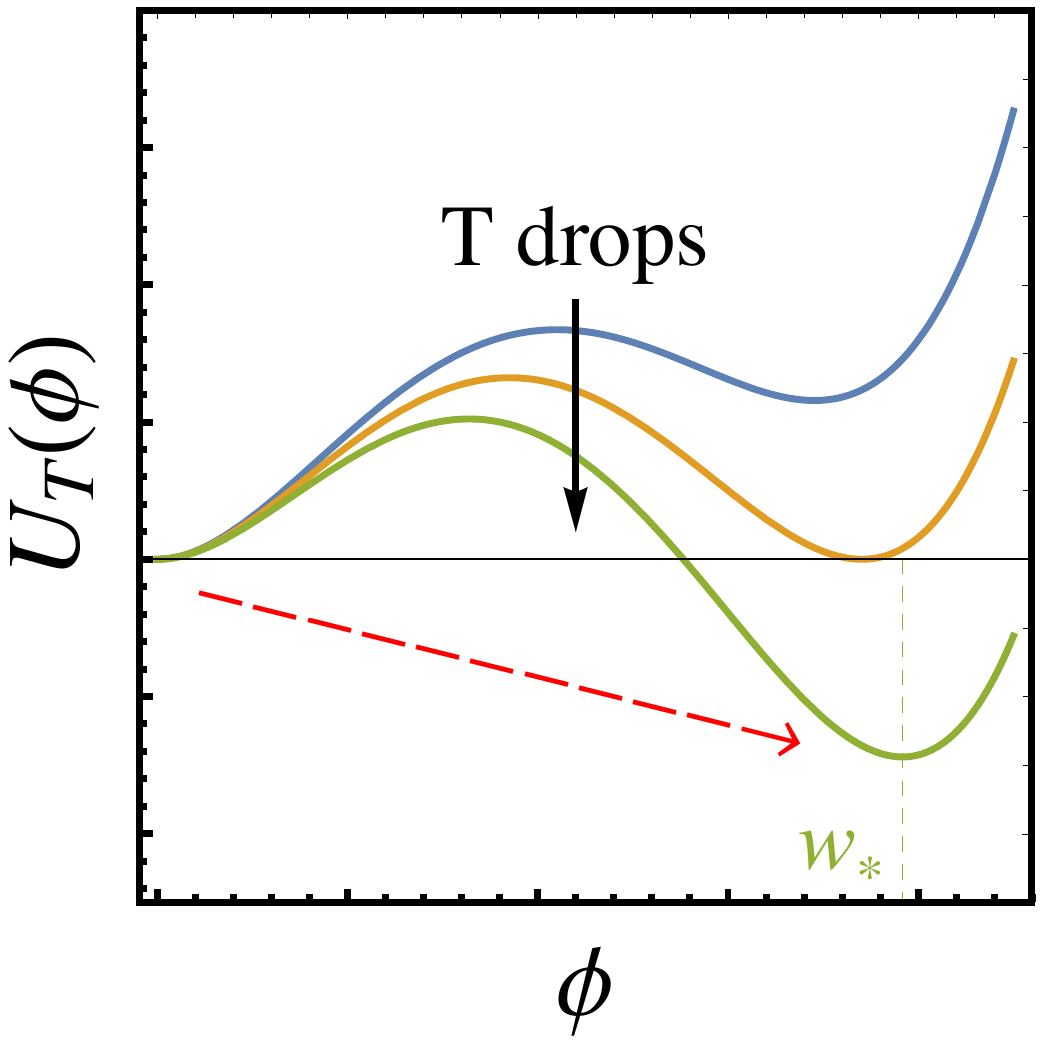}\qquad
\includegraphics[scale=0.273]{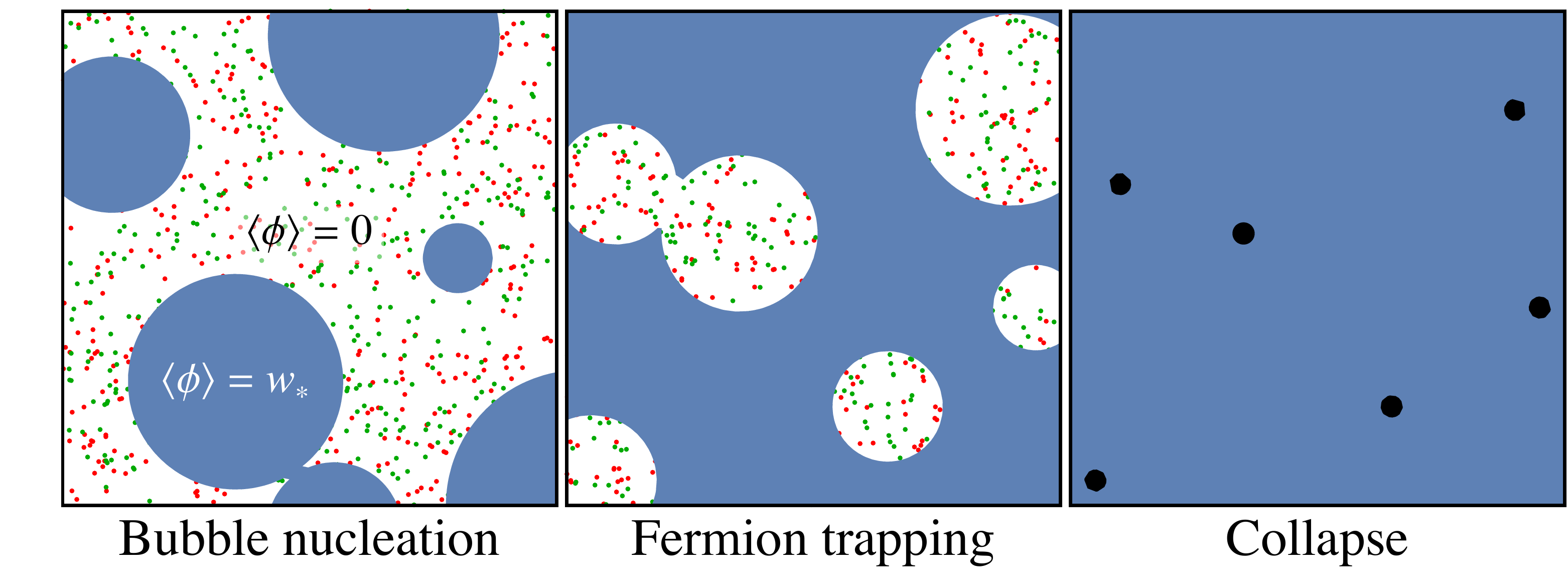}
\caption{{\bf Left}: in field space, a FOPT is the decay of the Universe between two vacua separated by a barrier. {\bf Right}: in spacetime, a FOPT is the nucleation and growth of vacuum bubbles (blue regions). If fermions (the red and green dots) are trapped in false vacuum remnants (white regions) and squeezed, those remnants might collapse into PBHs (black bold dots). See the text for details.}
\label{fig:FOPT}
\end{figure}

Suppose the Universe is filled up with a scalar field $\phi$, whose effective potential $U_T(\phi)$ evolves with the temperature $T$~\cite{Quiros:1999jp}. The vacuum is initially at the field space origin $\phi=0$, where the Universe stays at. As $T$ drops, the potential develops another deeper local minimum, or say, ``the true vacuum'', away from the origin. If the two vacua are separated by a potential barrier, the Universe cannot smoothly shift to the true vacuum, but can only decay to it through quantum tunneling~\cite{Linde:1981zj}, as illustrated in the left panel of Fig.~\ref{fig:FOPT}. This is known as a FOPT, which happens in spacetime via vacuum bubble nucleation and expansion. Inside the bubble is the new true vacuum with $\phi\neq0$, while outside the bubble is the old false vacuum with $\phi=0$. The FOPT completes when the bubbles eventually fulfill the entire space, converting the whole Universe to the true vacuum.

If there is a fermion species $\chi$ that couples to the scalar field $\phi$ via $\mL\supset\bar\chi i\slashed{\partial}\chi-y_\chi\phi\bar\chi\chi$, then during the FOPT $\chi$ would be massless outside the bubble, but gain a mass $m_\chi=y_\chi w_*$ inside the bubble, where $w_*$ is the vacuum expectation value (VEV) of $\phi$ at the true vacuum at FOPT temperature $T_*$. If the mass gap $m_\chi\gg T_*$, then most of the fermions are not able to penetrate into the true vacuum because the kinetic energy is $\mO(T_*)$~\cite{Baker:2019ndr,Chway:2019kft,Chao:2020adk,Deng:2020dnf}. For example, $m_\chi=12\,T_*$ yields a trapping fraction of 98\% when the bubble velocity is $v_w=0.6$~\cite{Kawana:2021tde,Hong:2020est}. As a result, the fermions are reflected by the bubble walls and hence get trapped in the false vacuum. One can naturally expect that, as the FOPT proceeds, the false vacuum remnants shrink to smaller and smaller sizes, and then the trapped fermions are squeezed, which means the energy density increases rapidly. Those remnants might eventually collapse into PBHs, as sketched in the right panel of Fig.~\ref{fig:FOPT}.

However, for the false vacuum remnants to be sufficiently dense to collapse into PBHs, we have to address an important issue: how to prevent the trapped fermions from disappearing through $\chi\bar\chi\to\phi,~\phi\phi$, etc, and eventually to the SM particles in the thermal bath? Such annihilations are inevitably enhanced when the remnants shrink, reducing the fermion number density greatly~\cite{Arakawa:2021wgz,Asadi:2021yml} and consequently destroying any possibility of collapse into PBHs. To have PBHs formed, there are two typical scenarios:
\begin{enumerate}
\item[I.] Suppress the annihilation rate by a relatively small Yukawa coupling $y_\chi$~\cite{Baker:2021nyl,Baker:2021sno}. In this case, the false vacuum remnants directly collapse into PBHs.
\item[II.] Generate a $\chi$-$\bar\chi$ number density asymmetry such that $\chi$'s survive the annihilation~\cite{Kawana:2021tde}. In this case, the remnants first shrink to non-topological solitons called ``Fermi-balls''~\cite{Hong:2020est}, which could collapse into PBHs due to the internal Yukawa interaction.
\end{enumerate}
We will discuss them one by one in the following subsections. In those scenarios, the distribution of false vacuum remnants is crucial in deriving the PBH mass function. While the numerical study of such a distribution is still lacking, we adopt the analytical technique developed in Ref.~\cite{Lu:2022paj} as a first trial.

\subsection{Scenario I: the direct collapse of false vacuum remnants}\label{subsec:2105PBH}

This scenario is first proposed by Refs.~\cite{Baker:2021nyl,Baker:2021sno}, which demonstrate that by adopting a small Yukawa coupling 
\be\label{smally}
y_\chi\lesssim10^{-4}\left(\frac{T_*}{10^6~{\rm GeV}}\right)^{1/2},
\ee
the annihilation processes $\chi\bar\chi\to\phi,~\phi\phi$ are suppressed. Note that the trapping condition $m_\chi=y_\chi w_*\gg T_*$ requires $w_*\gg T_*$ when $y_\chi$ is small. Once \Eq{smally} holds, the energy density of the trapped fermions is approximately
\be\label{r4scaling}
\rho_\chi(t)\approx\rho_\chi^{\rm eq}\left(\frac{R_r^0}{R_r(t)}\right)^4,
\ee
where $R_r(t)$ is the size of the false vacuum remnant at time $t$, $R_r^0\equiv R_r(t=t_*)$ is the initial size with $t_*$ being the cosmic time of FOPT, $\rho_\chi^{\rm eq}=(7/8)(\pi^2g_\chi/30)T_*^4$ is the initial fermion energy density of the remnant (which is just the equilibrium value right before the FOPT), $g_\chi=4$ counts the number of the degrees of freedom including both $\chi$ and $\bar\chi$. \Eq{r4scaling} includes the number density enhancement $\propto\left(R_r^0/R_r(t)\right)^3$ and the energy gain of the fermions from reflections of the bubble wall $\propto R_r^0/R_r(t)$ during the shrinking of $R_r(t)$. The scaling in \Eq{r4scaling} is confirmed by the numerical simulations~\cite{Baker:2021nyl,Baker:2021sno} and the analytical calculations~\cite{Kawana:2022lba}.

As $R_r(t)$ decreases, $\rho_\chi(t)$ increases and so does the Schwarzschild radius of the remnant,
\be
R_s(t)=\frac{2}{M_{\rm Pl}^2}\frac{4\pi}{3}R_r^3(t)\rho_\chi(t),
\ee
with $M_{\rm Pl}=1.22\times10^{19}$ GeV the Planck scale. When $R_r(t)<R_s(t)$, the remnant collapses into a black hole. Denote the collapse time as $t_{\rm col}$, the collapse condition is then
\be\label{collapse2105}
\frac{R_r(t_{\rm col})}{R_r^0}=\sqrt{\frac78\frac{g_\chi}{g_*}}\frac{R_r^0}{H_*^{-1}},
\ee
with $H_*^{-1}$ being the Hubble radius at FOPT. Note that the collapse condition is determined by the overdense region of the $\chi/\bar\chi$ fermions only.

\Eq{collapse2105} implies that, at the moment of PBH formation, the ratio of the remnant size $R_r(t_{\rm col})$ to the initial size $R_r^0$ is proportional to the ratio of $R_r^0$ to the Hubble radius $H_*^{-1}$. As mentioned, the premise of this scenario is the suppressed $\chi\bar\chi$ annihilation, such that we have \Eq{r4scaling} and hence \Eq{collapse2105}; however, the annihilation is extremely difficult to suppress while $R_r(t_{\rm col})/R_r^0$ is too small. Therefore, we can infer that, in this direct collapse scenario, the PBH formation prefers an initial condition with large $R_r^0/H_*^{-1}$. Indeed, the simulations in Refs.~\cite{Baker:2021nyl,Baker:2021sno} show successful examples of PBH formation for $R_r^0=1.5H_*^{-1}$ and $2H_*^{-1}$. The resultant PBH mass from the collapse is estimated by the total energy contained in the false vacuum remnant,
\be\label{mpbh2105}
M\approx\frac{4\pi}{3}R_r^3(t_{\rm col})\left(\frac{3M_{\rm Pl}^2}{8\pi}H_*^2\right)=\left(\frac{7g_\chi}{8g_*}\right)^{3/2}\frac{M_{\rm Pl}^2}{2H_*}\left(\frac{R_r^0}{H_*^{-1}}\right)^6,
\ee
and we assume the PBH formation is possible only for $R_r^0>H_*^{-1}$ for simplicity.

According to \Eq{mpbh2105}, to get the mass distribution of $M$, we should first derive the distribution of the false vacuum remnant size $R_r^0$ during a FOPT. This can be done using the method proposed by Ref.~\cite{Lu:2022paj},
\be\label{dndRr}
\frac{\d n_{\rm fv}}{\d R_r^0}\approx\frac{I_*^4\beta^4}{192v_w^3}e^{(4\beta R_r^0/v_w)-I_*e^{\beta R_r^0/v_w}}
\left(1-e^{-I_*e^{\beta R_r^0/v_w}}\right)~,
\ee
where $n_{\rm fv}$ is the number density of the remnants during the FOPT, $I_\ast=-\ln(0.29)= 1.238$, and $\beta$ is the reciprocal of the FOPT duration. Each remnant with radius $R_r^0$ larger than $H_*^{-1}$ collapses into one individual PBH, thus the number density distribution of PBHs at formation is
\be
\frac{\d n_{\rm pbh}^*}{\d M}=\frac{\d n_{\rm fv}}{\d R_r^0}\Big|_{R_r^0>H_*^{-1}}\left(\frac{\d M}{\d R_r^0}\right)^{-1},
\ee
and current PBH distribution should be
\be\label{dfdm2105}
\frac{\d f_{\rm pbh}}{\d M}=\left(\frac{M}{M'}\right)^3\frac{1}{\Omega_{\rm DM}}\left(\frac{8\pi}{3M_{\rm Pl}^2H_0^2}\right)\frac{s_0}{s_*}\left(M\frac{\d n_{\rm pbh}^*}{\d M}\right)\Big|_{M\to M'},
\ee
where $M'=(M^3+3f_0M_{\rm Pl}^4t_0)^{1/3}$ accounts for the mass loss of PBHs after formation, $s_0=2891.2~{\rm cm}^{-3}$ is the current entropy density, and $H_0=67.4~{\rm km/(s\cdot Mpc)}$ is the current Hubble constant~\cite{ParticleDataGroup:2020ssz}.

\begin{figure}
\centering
\includegraphics[scale=0.39]{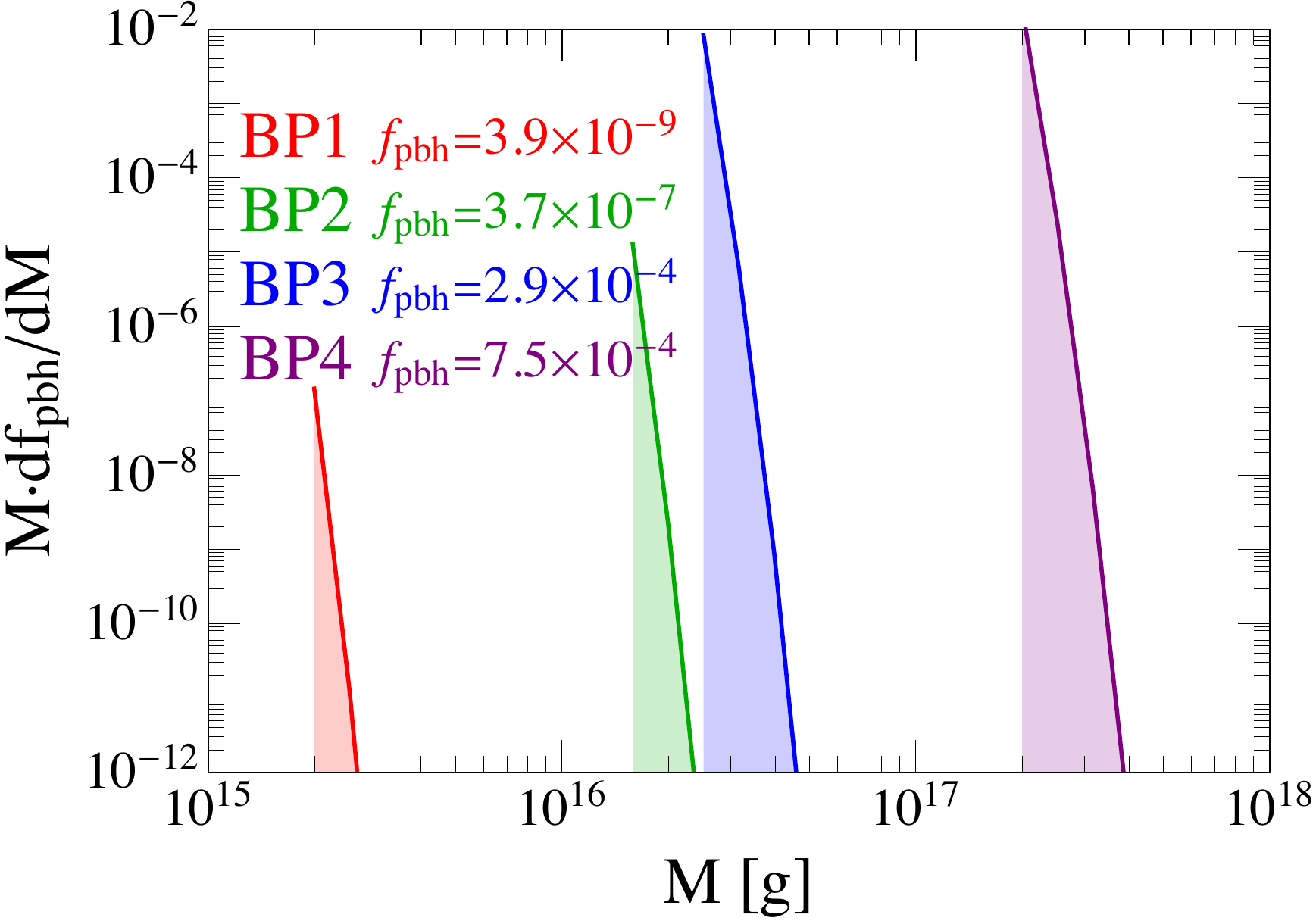}\qquad
\includegraphics[scale=0.4]{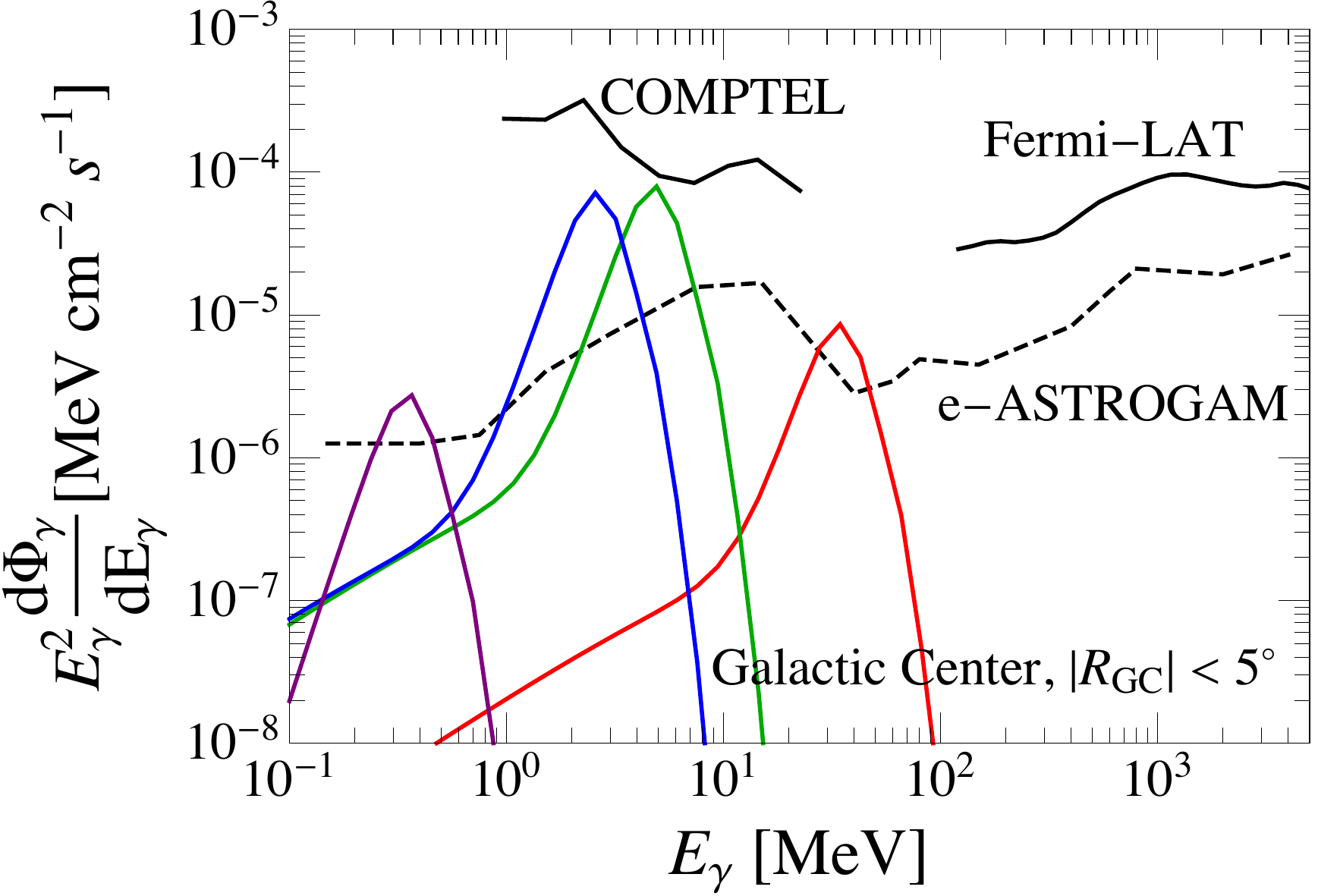}
\caption{The current mass distributions ({\bf left}) and corresponding gamma-ray spectra ({\bf right}) of the FOPT-induced PBHs in the direct collapse scenario. The description of the BPs can be found in \Eq{BPs_2105}.}
\label{fig:2105}
\end{figure}

Eqs.~(\ref{mpbh2105})--(\ref{dfdm2105}) are used to derive the PBH mass function, and the input parameters are $\{\alpha,~\beta/H_*,~T_*,~v_w\}$. As PBHs form only when $R_r^0>H_*^{-1}$, while $\d n_{\rm fv}/\d R_r^0$ decreases rapidly with $R_r^0$ due to the double-exponential suppression factor, we can expect $\d f_{\rm pbh}/\d M$ has a very sharp peak at around
\be
M_{\rm peak}\approx\left(\frac{7g_\chi}{8g_*}\right)^{3/2}\frac{M_{\rm Pl}^2}{2H_*}=5.2\times10^{15}~{\rm g}\times\left(\frac{10^7~{\rm GeV}}{T_*}\right)^2,
\ee
where $g_*=g_\chi+106.75$ is adopted in the last line. Due to this reason, the gamma-ray signal shape is also very narrow. We perform a parameter scan for $\alpha\in[0,1]$, $\beta/H_*\in[1,10^3]$, $T_*\in[10^{-2},10^{8}]~{\rm GeV}$ and $v_w\in[0.1,0.8]$, requiring that $M_{\rm peak}$ is in the asteroid-mass range, the gamma-ray signals are allowed by current observations, and $f_{\rm pbh}\leqslant1$. The requirement of mass leads to $T_*\sim10^7$ GeV, as expected. The gamma-ray signals have narrow shapes at MeV scale, as can be obtained clearly in Fig.~\ref{fig:2105}, where we plot the corresponding mass and gamma-ray spectra in the left and right panels respectively for the four BPs
\be\label{BPs_2105}
\{\alpha,~\beta/H_*,~T_*,~v_w\}=\begin{cases}~\{0.0827,~2.21,~1.65\times10^7,~0.566\},&{\rm BP1};\\
~\{0.202,~1.93,~6.22\times10^6,~0.514\},&{\rm BP2};\\
~\{0.788,~2.90,~4.58\times10^6,~0.781\},&{\rm BP3};\\
~\{0.308,~1.83,~1.68\times10^6,~0.505\},&{\rm BP4},\\
\end{cases}
\ee
randomly chosen from the scanned results.

\begin{figure}
\centering
\includegraphics[scale=0.4]{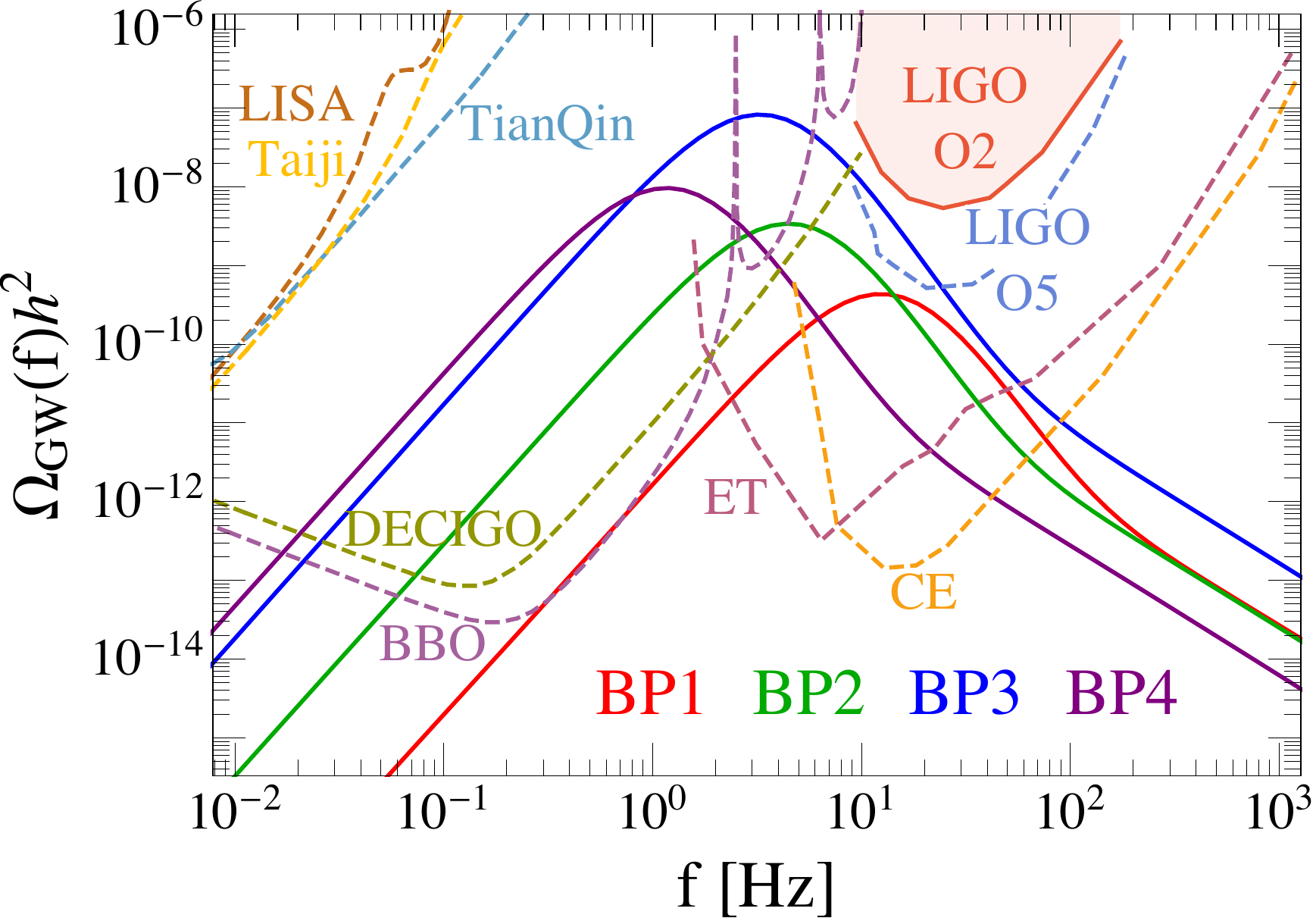}
\caption{The FOPT GW spectra of the BPs in \Eq{BPs_2105}, which are correlated signals from PBHs from direct collapse during a FOPT.}
\label{fig:gw_2105}
\end{figure}

FOPT generates stochastic GWs via bubble collisions, sound waves, and turbulences. The resultant GW spectrum, after the cosmological redshift, can be expressed as a function of the FOPT parameters $\{\alpha,\beta/H_*,T_*,v_w\}$, namely the ratio of latent heat to the radiation energy density, the inverse ratio of transition duration to the Hubble time, the FOPT temperature and the bubble expansion velocity~\cite{Caprini:2015zlo,Caprini:2019egz}. For particle trapping scenarios, $v_w$ is not extremely close to 1 and hence the dominant contribution to GW spectrum is from the sound waves in the plasma, which yields a peak at~\cite{Caprini:2015zlo}\footnote{The amplitude of the GW signal may be suppressed by the finite duration of the sound wave period~\cite{Ellis:2018mja,Guo:2020grp}. However, in the parameter space of interest, the FOPT duration is typically long that $\beta/H_*\lesssim10$, and hence the sound wave suppression is not prominent.}
\be\label{gw2105}\begin{split}
f_{\rm peak}\approx&~19~{\rm Hz}\times\left(\frac{0.1}{v_w}\right)\left(\frac{\beta}{H_*}\right)\left(\frac{T_*}{10^7~{\rm GeV}}\right)\left(\frac{g_*}{100}\right)^{1/6},\\
\Omega_{\rm sw}(f_{\rm peak})h^2\approx&~2.65\times10^{-7}\times\left(\frac{\beta}{H_*}\right)^{-1}\left(\frac{\kappa_V\alpha}{1+\alpha}\right)^2\left(\frac{g_*}{100}\right)^{-1/3}\left(\frac{v_w}{0.1}\right),
\end{split}\ee
where $\kappa_V$ is the fraction of released vacuum energy that goes into the plasma kinetic bulk motion~\cite{Espinosa:2010hh}. The corresponding GW signals are within the sensitive region of the ground-based detectors LIGO~\cite{LIGOScientific:2014qfs,LIGOScientific:2019vic}, CE~\cite{Reitze:2019iox} and ET~\cite{Punturo:2010zz,Hild:2010id,Sathyaprakash:2012jk}, or the space-based detectors BBO~\cite{Crowder:2005nr} and DECIGO~\cite{Kawamura:2011zz}. For the four BPs in \Eq{BPs_2105}, we plot the GW spectra in Fig.~\ref{fig:gw_2105}. We have checked that BPs show the representative features (e.g. frequencies, signal strengths) of the scanned parameter points.

Here we provide a short remark for the direct collapse scenario of the FOPT-induced PBHs. Requiring a large false vacuum remnant size $R_r^0$, the PBH mass function in this scenario features a sharp peak. As a result, the gamma-ray spectrum also has a very narrow distribution, as illustrated in Fig.~\ref{fig:2105}. The correlated FOPT GW signals are expected to peak at around $\mO(10)$ Hz, as shown in Fig.~\ref{fig:gw_2105}.

\subsection{Scenario II: collapse of FOPT-induced solitons}\label{subsec:solitons}

This scenario is first proposed by Ref.~\cite{Kawana:2021tde} and then applied in Refs.~\cite{Marfatia:2021hcp,Huang:2022him,Tseng:2022jta,Kawana:2022lba,Lu:2022jnp,Marfatia:2022jiz}. A baryogenesis-like mechanism (or say, asymmetric DM~\cite{Kaplan:2009ag,Petraki:2013wwa,Zurek:2013wia}) is introduced to have $n_\chi>n_{\bar\chi}$ before or during the FOPT, and hence when the fermions are trapped in the false vacuum remnants and forced to annihilate, $\bar\chi$'s will disappear eventually, but $\chi$'s survive. Those residual fermions develop a degeneracy pressure when they are compressed. When that pressure is able to balance the vacuum pressure, a non-topological soliton solution exists, known as the Fermi-ball~\cite{Hong:2020est}. This happens at $\d E_{\rm rem}/\d R_r=0$ for the following remnant energy profile
\be\label{FB_E}
E_{\rm rem}\approx\frac{3\pi}{4}\left(\frac{3}{2\pi}\right)^{2/3}\frac{Q_{\rm FB}^{4/3}}{R_r}+\frac{4\pi}{3}\Delta U(T_*)R_r^3,
\ee
where $R_r$ is the remnant radius, $Q_{\rm FB}$ is the charge, i.e. the number of residual fermions trapped in an individual remnant, and $\Delta U(T_*)$ is the positive vacuum energy difference between the true and false vacua that represents the vacuum pressure that drives the expansion of the bubbles. The Fermi-ball mass and radius are respectively~\cite{Hong:2020est}\footnote{Trapping particles to form non-topological soliton is an old idea that receives renewed interest recently. See Refs.~\cite{Krylov:2013qe,Huang:2017kzu,Bai:2022kxq} for the Q-balls from a FOPT, and Refs.~\cite{Witten:1984rs,Frieman:1990nh,Zhitnitsky:2002qa,Oaknin:2003uv,Lawson:2012zu,Atreya:2014sca,Bai:2018vik,Bai:2018dxf,Gross:2021qgx} for the quark nuggets or dark dwarfs from a QCD-like FOPT.}
\be\label{MR_profile}
M_{\rm FB}\approx Q_{\rm FB}\left(12\pi^2\Delta U(T_*)\right)^{1/4},\quad
R_{\rm FB}^3\approx\frac{3}{16\pi}\frac{M_{\rm FB}}{\Delta U(T_*)},
\ee
dominated by the charge and vacuum energy difference. Fermi-balls are very dense objects, and they could collapse into PBHs if the internal $\phi$-mediated Yukawa attractive force between the fermions is too strong~\cite{Kawana:2021tde}. In that case, the daughter PBH inherits the mother Fermi-ball's mass, $M\approx M_{\rm FB}$.

The charge of a given Fermi-ball could be expressed as~\cite{Kawana:2021tde}
\be\label{QFB}
Q_{\rm FB}=F_\chi^{\rm trap.}\frac{\eta_\chi s_*}{p_*}\frac{4\pi}{3}\left(R_r^0\right)^3,
\ee
where $F_\chi^{\rm trap.}\approx1$ is the fermion trapping fraction, $\eta_\chi=(n_\chi-n_{\bar\chi})/s$ is the $\chi$-asymmetry, $s_*$ is the entropy density at FOPT, $R_r^0$ is the initial radius of the false vacuum remnant, and $p_*=0.29$ based on percolation condition~\cite{Hong:2020est}. Combining \Eq{QFB} with \Eq{MR_profile}, one finds that $M$ this scenario is also determined by the $R_r^0$ distribution, which is given by \Eq{dndRr}. The other parameter, vacuum energy, can be parameterized as $\Delta U(T_*)\approx(\pi^2/30)g_*T_*^4\alpha$, and hence the PBH mass distribution $\d f_{\rm pbh}/\d M$ can be expressed as a function of FOPT parameters $\{\alpha, \beta/H_*, T_*, v_w\}$ following similar logics described in Section~\ref{subsec:2105PBH}.

\begin{figure}
\centering
\includegraphics[scale=0.39]{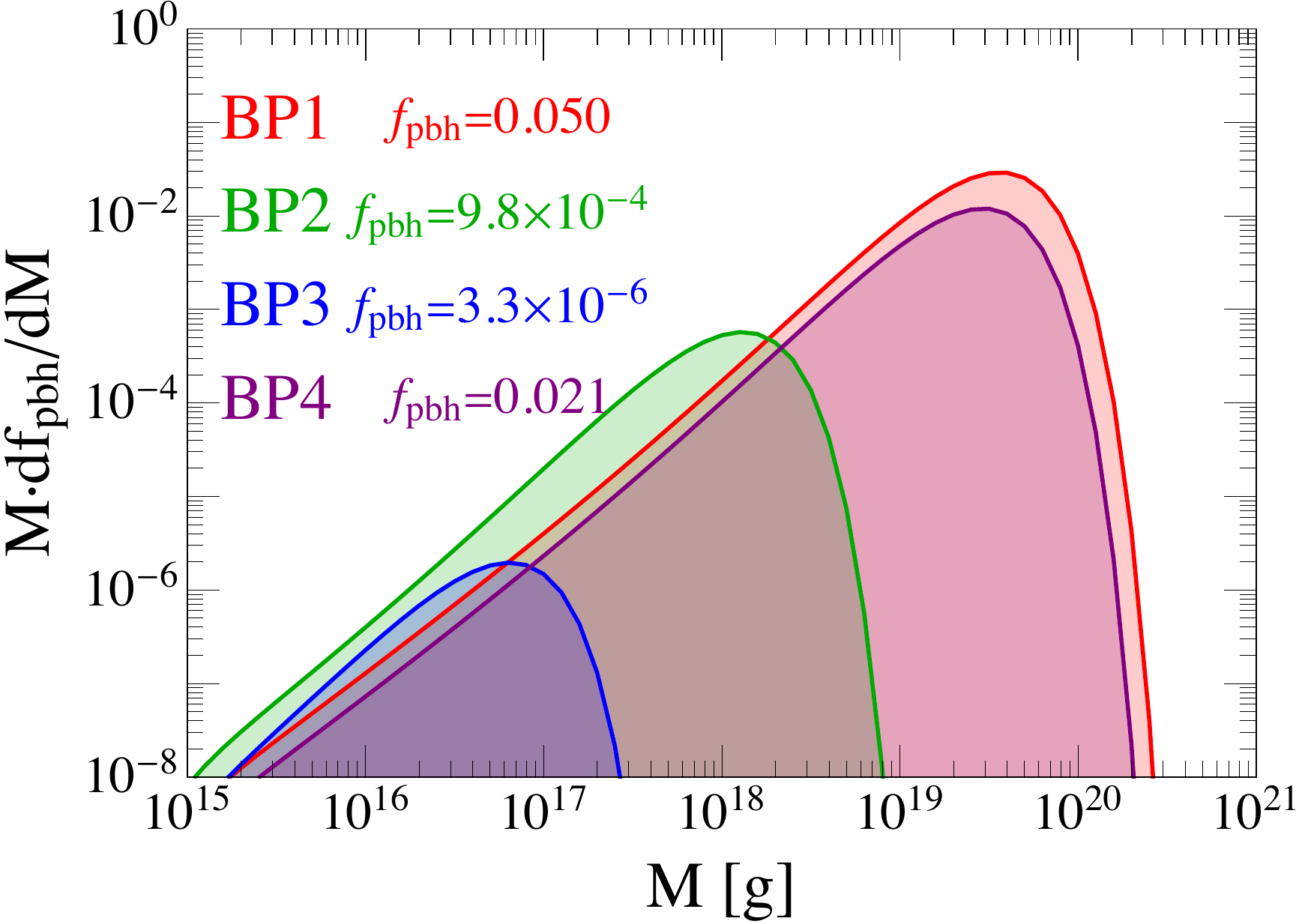}\qquad
\includegraphics[scale=0.4]{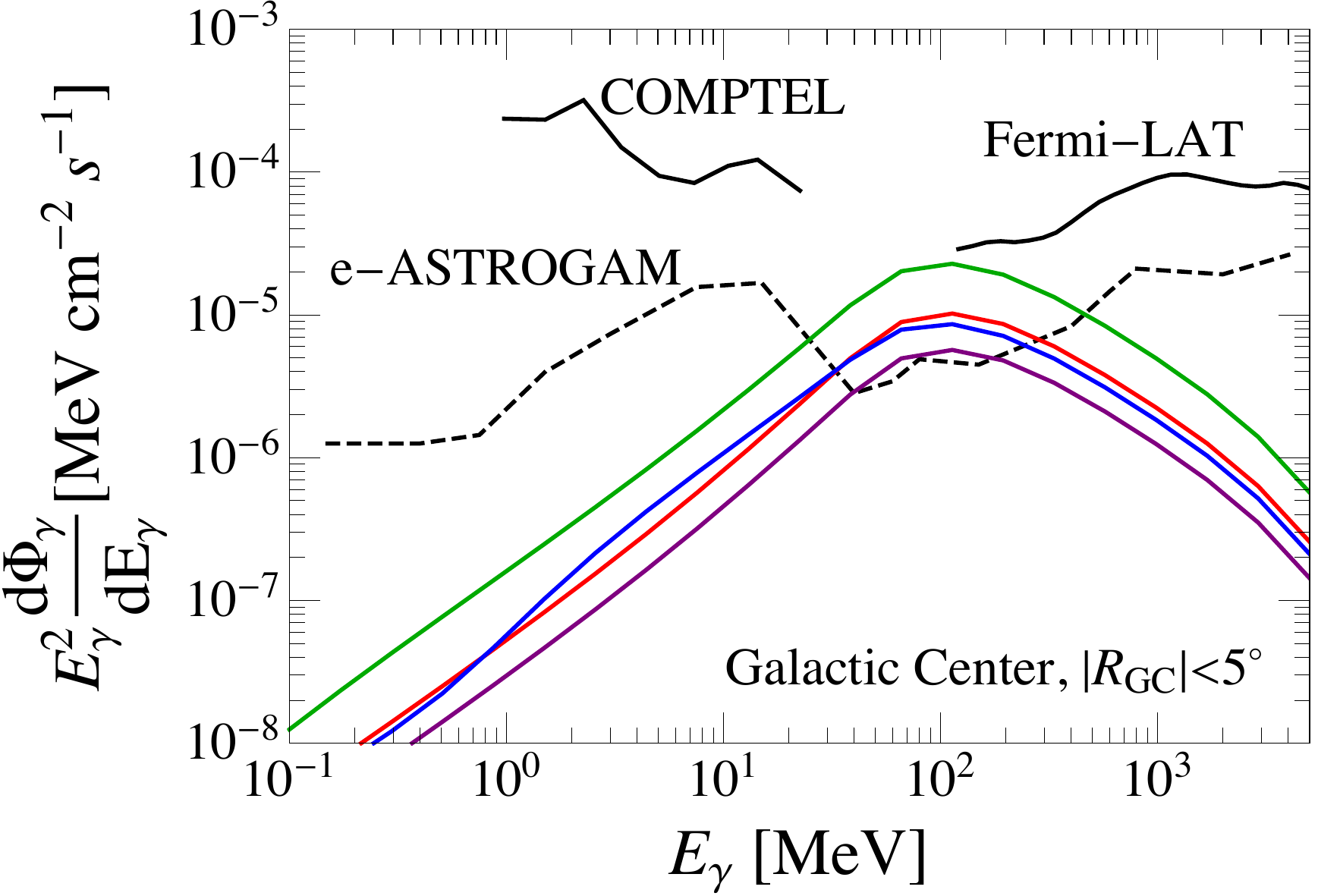}
\caption{The current mass distributions ({\bf left}) and corresponding gamma-ray spectra ({\bf right}) of the FOPT-induced PBHs in the Fermi-ball collapse scenario. The description of the BPs can be found in \Eq{BPs_kpx}.}
\label{fig:kpx}
\end{figure}

We can see $M\propto(R_r^0)^3$. Since the distribution of $R_r^0$ has a peak around $v_w/\beta$~\cite{Lu:2022paj}, $\d f_{\rm pbh}/\d M$ also has a peak, which can be estimated as~\cite{Kawana:2021tde}
\be\label{m_2106}
M_{\rm peak}\approx 1.4\times10^{16}~{\rm g}\times \left(\frac{v_w}{0.1}\right)^3\left(\frac{\eta_\chi}{10^{-12}}\right)
\left(\frac{100}{g_*}\right)^{1/4}\left(\frac{\rm GeV}{T_*}\right)^2\left(\frac{10}{\beta/H_*}\right)^3\alpha^{1/4}~.
\ee
Notice that there are many variables affecting the peak position of $M$. To insure the generality, we scan over $\alpha\in[0,1]$, $\beta/H_*\in[1,10^3]$, $T_*\in[10^{-2},10^8]$ GeV, $v_w\in[0.1,0.8]$, and $\eta_\chi\in[10^{-3},10^{-15}]$. By requiring the derived PBH mass functions to peak within the asteroid-mass range, to escape the current gamma-ray or other bounds, and to satisfy $f_{\rm pbh}\leqslant1$, we obtain the typical parameter space, which is reflected in the normalized numbers in \Eq{m_2106}. Four BPs are chosen as 
\be\label{BPs_kpx}
\{\alpha,~\beta/H_*,~T_*,~v_w,~\eta_\chi\}=\begin{cases}~\{0.988,~3.91,~2.48,~0.777,~1.51\times10^{-12}\},&{\rm BP1};\\
~\{0.753,~25.1,~0.334,~0.797,~2.43\times10^{-13}\},&{\rm BP2};\\
~\{0.326,~3.92,~0.407,~0.252,~2.64\times10^{-15}\},&{\rm BP3};\\
~\{0.199,~4.26,~1.79,~0.754,~1.33\times10^{-12}\},&{\rm BP4},\\
\end{cases}
\ee
to plot the mass functions and gamma-ray spectra in Fig.~\ref{fig:kpx}. The shape of $\d f_{\rm pbh}/\d M$ is milder than the direct collapse scenario, and hence the gamma-ray spectra are much broader.

\begin{figure}
\centering
\includegraphics[scale=0.4]{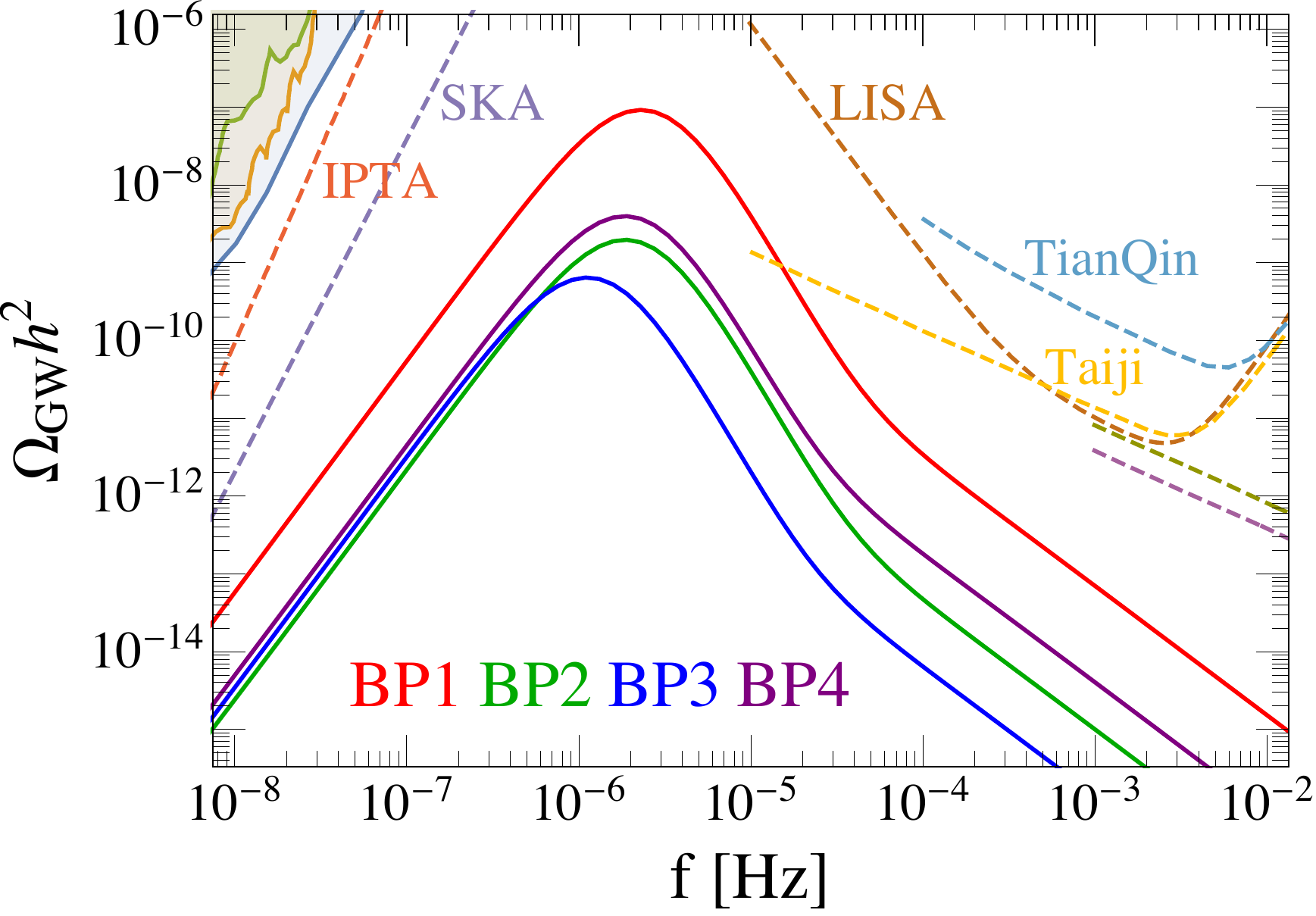}
\caption{The FOPT GW spectra of the BPs in \Eq{BPs_kpx}, which are correlated signals from PBHs from collapse of Fermi-balls formed in a FOPT.}
\label{fig:gw_kpx}
\end{figure}

The accompanied GW signals can be calculated by $\{\alpha,\beta/H_*,T_*,v_w\}$, as stated before~\cite{Caprini:2015zlo,Caprini:2019egz}. In this scenario, the GWs peak at
\be
f_{\rm peak}\approx1.9\times10^{-6}~{\rm Hz}\times\left(\frac{0.1}{v_w}\right)\left(\frac{\beta}{H_*}\right)\left(\frac{T_*}{{\rm GeV}}\right)\left(\frac{g_*}{100}\right)^{1/6},
\ee
where the normalized numbers are chosen according to the parameter scan described above. Unfortunately, this frequency region is not reachable by any current or future GW detectors. As illustrated in Fig.~\ref{fig:gw_kpx}, the GW signals of the four BPs lie between the sensitive regions of the PTAs and the space-based interferometers, and this is also the general feature of the whole allowed parameter space. It is proposed that GWs with frequencies $\mu{\rm Hz}$ can be detected by future space-based $\mu$Ares~\cite{Sesana:2019vho} or asteroid-based~\cite{Fedderke:2021kuy} interferometers, but there is still a long way to go to the final realization of those ideas. Therefore, the characteristic signal of the Fermi-ball collapse PBHs is that we can only see a mild gamma-ray spectrum, with no detected GWs. However, as implied by \Eq{m_2106}, the asteroid-mass PBHs in this mechanism favor a FOPT at GeV scale, which might be probed by BBN and CMB observations~\cite{Bai:2021ibt,Liu:2022lvz} and show some correlation features with the PBH detection.

\section{PBHs from cosmic strings}\label{sec:cs}

The spontaneous breaking of a $U(1)$ symmetry could form one-dimensional topological defects, known as the cosmic strings~\cite{Nielsen:1973cs,Kibble:1976sj,Vachaspati:2015cma}. When the symmetry is broken at a high scale $w$, long string networks with a tension $\mu\sim w^2$ are formed. Those long strings are relativistic objects interacting and colliding frequently with themselves or each other, which continuously produces sub-horizon small string loops. The small loops then keep emitting GWs and shrinking until they disappear, generating the stochastic GW background today~\cite{Auclair:2019wcv}. A sub-horizon string may collapse into a PBH, if it shrinks to a size smaller than its Schwarzschild radius~\cite{Hawking:1987bn,Polnarev:1988dh}. The fraction of cosmic strings that collapse into PBHs can be constrained by the gamma-rays from Hawking radiation~\cite{Caldwell:1993kv,MacGibbon:1997pu} or CMB distortions~\cite{James-Turner:2019ssu,Bianchini:2022dqh}.

Here we adopt the calculations in Refs.~\cite{Caldwell:1993kv,MacGibbon:1997pu,James-Turner:2019ssu,Caldwell:1991jj} to derive the PBH mass function from cosmic string collapse. The energy density of the small string loops, $\rho_\ell$, evolve as
\be\label{loops_evolve}
\dot\rho_\ell+3H\rho_\ell=-\left[\dot\rho_\infty+2H\rho_\infty\left(1+\langle v^2\rangle\right)\right]\delta,
\ee
where $H=1/(2t)$ is the Hubble constant, $\rho_\infty$ is the energy density of long string networks, and numerical simulations show $\langle v^2\rangle\approx0.4$~\cite{Blanco-Pillado:2011egf} and $\delta\approx0.1$~\cite{Blanco-Pillado:2013qja}. The physical meaning of \Eq{loops_evolve} is clear: small loops are being chopped off from the long strings, and hence a fraction of energy is transferred from the network system to the loop system. After reaching the scaling regime, $\rho_\infty\approx A\mu/(4t^2)$ with $A\approx44$~\cite{Blanco-Pillado:2011egf,Blanco-Pillado:2019vcs}. Therefore, the right-hand side of \Eq{loops_evolve} is known. As for the left-hand side, $\rho_\ell=n_\ell\times\mu R_\ell$, where $n_\ell$ is the number density of loops, $R_\ell$ is the length of an individual loop, which we assume to be a fixed fraction $\alpha_\ell\approx0.1$ of the Hubble radius $H^{-1}=2t$, i.e. $R_\ell=2\alpha_\ell t$~\cite{Blanco-Pillado:2011egf}. If we further assume a fraction of $\epsilon$ of the small loops collapse into PBHs, then the number density of PBHs $n_{\rm pbh}=\epsilon\, n_\ell$. Combining all information above, one obtains the evolution of PBH number density as
\be\label{npbhcs}
\dot n_{\rm pbh}+3Hn_{\rm pbh}=\epsilon\frac{A\delta}{8\alpha_\ell t^4}\left(1-\langle v^2\rangle\right).
\ee
The mass of PBH at time $t$ is $M\approx\mu\alpha_\ell t$.

\begin{figure}
\centering
\includegraphics[scale=0.39]{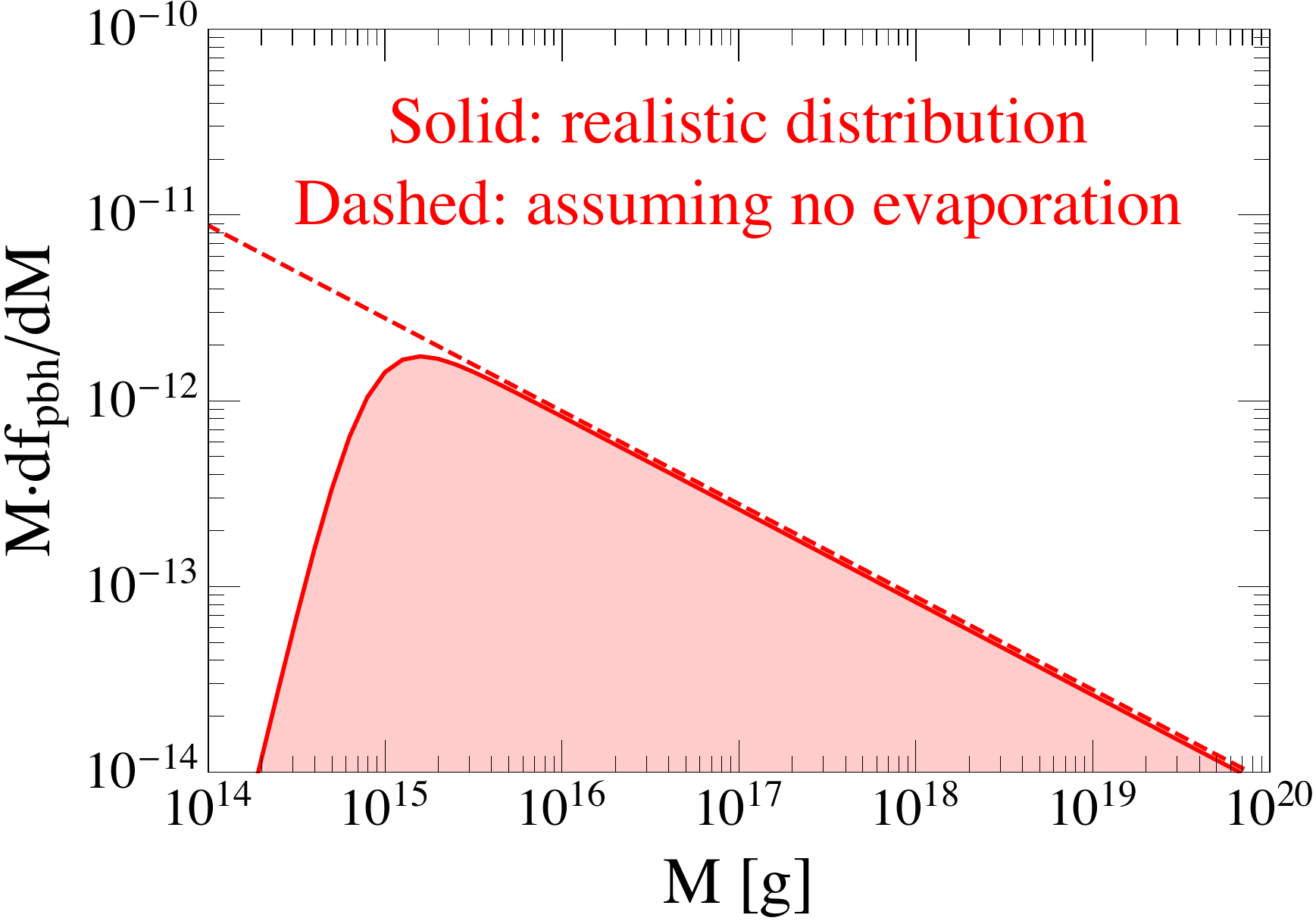}\qquad
\includegraphics[scale=0.4]{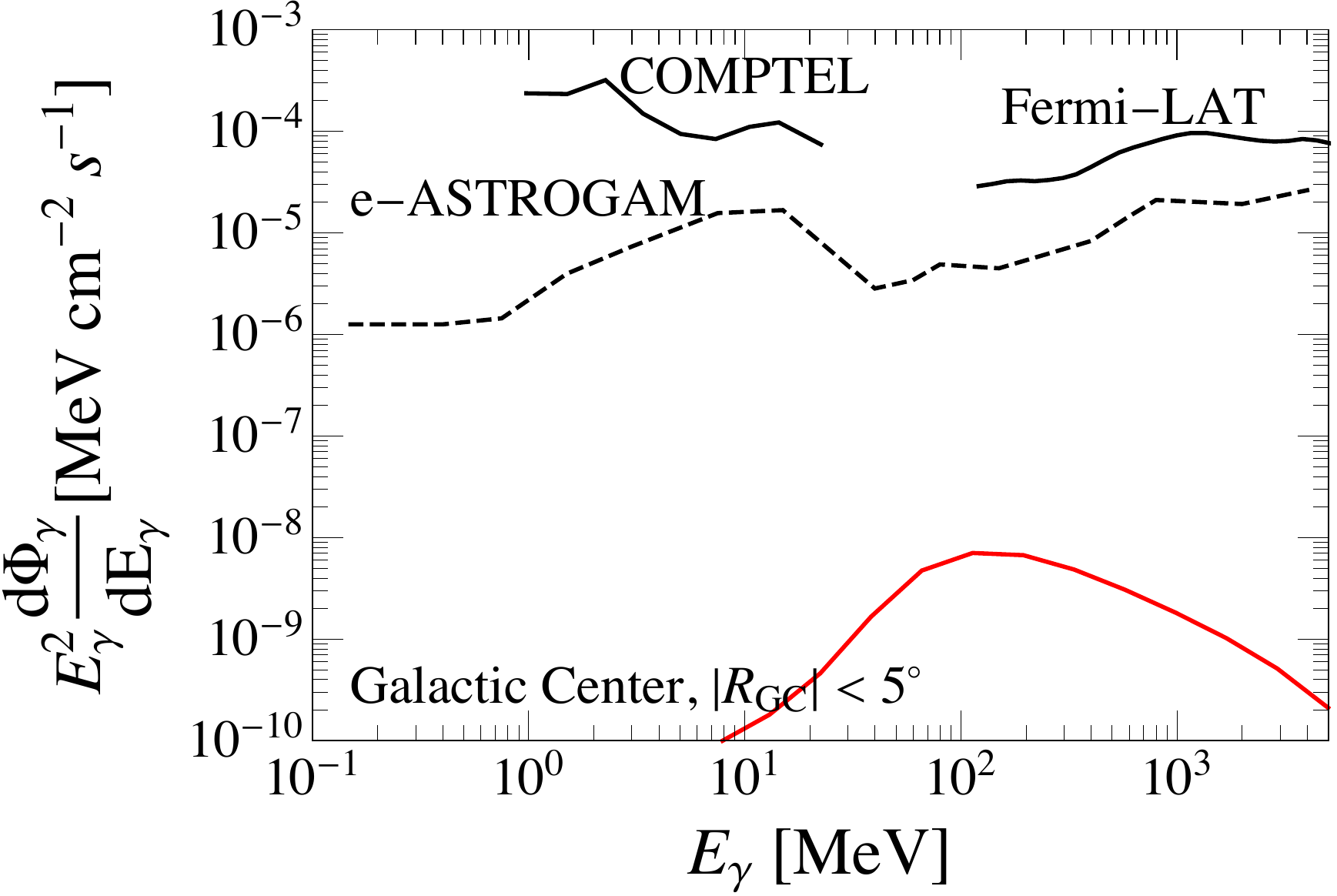}
\caption{The current mass distributions ({\bf left}) and corresponding gamma-ray spectra ({\bf right}) of the PBH from cosmic string collapse, for the maximal $\epsilon\,\mu^{3/2}$ allowed by the current experiments.}
\label{fig:cs}
\end{figure}

The PBHs from cosmic strings are mainly produced deep in the radiation era, as implied by \Eq{npbhcs}. Integrating this equation up to the matter-radiation equality, we find
\be\label{pbh_cs_eq}
\frac{\d f_{\rm pbh}}{\d M}\Big|_{\rm eq}=\frac{\Omega_m}{\Omega_{\rm DM}}\left(\epsilon\frac{\alpha_\ell^{1/2}\mu^{3/2}A\delta}{8t_{\rm eq}^{3/2}}\frac{1-\langle v^2\rangle}{M^{3/2}}\right)\Big/\left(\frac{\pi^2}{30}g_*^{\rm eq}T_{\rm eq}^4\right).
\ee
taking into account the PBH evaporation, the mass function today is
\be
\frac{\d f_{\rm pbh}}{\d M}=\left(\frac{M}{M'}\right)^3\frac{\d f_{\rm pbh}^{\rm eq}}{\d M}\Big|_{M\to M'},
\ee
where $M'=(M^3+3f_0M_{\rm Pl}^4t_0)^{1/3}$. Although \Eq{pbh_cs_eq} shows a $M^{-3/2}$ scaling, the evaporation effect makes current $\d f_{\rm pbh}/\d M\to0$ for $M\lesssim 5\times10^{14}$ g, as illustrated in the left panel of Fig.~\ref{fig:cs}.

On the other hand, even though having evaporated, the light PBHs still leave impacts in the CMB. Ref.~\cite{James-Turner:2019ssu} reveals that the CMB anisotropies have constrained
\be\label{cs_constraint}
\epsilon\lesssim10^{-8}\left(\frac{0.1}{\delta}\right)\left(\frac{44}{A}\right)\left(\frac{0.1}{\alpha_\ell}\right)^{1/2}\left(\frac{0.6}{1-\langle v^2\rangle}\right)\left(\frac{10^{-15}}{G\mu}\right)^{3/2},
\ee
where
\be
G\mu\equiv\frac{\mu}{M_{\rm Pl}^2}\sim10^{-15}\left(\frac{w}{4\times10^{11}~{\rm GeV}}\right)^2.
\ee
Note that \Eq{cs_constraint} has set a constraint on $\epsilon\,\mu^{3/2}$, which is also an overall factor of $\d f_{\rm pbh}/\d M$, see \Eq{pbh_cs_eq}. Since $\epsilon$ and $\mu$ are the only two free parameters in our simplified model, if we adopt the maximally allowed $\epsilon$ for a given $\mu$, then $\d f_{\rm pbh}/\d M$ is already fixed. This can be treated as the largest cosmic string-induced PBH abundance allowed by current experiments.

We then evaluate the gamma-ray spectrum of the above PBH mass function. Unfortunately, as shown in the right panel of Fig.~\ref{fig:cs}, it turns out that the gamma-ray signals are so weak that even future detectors cannot probe them. This conclusion is robust against the variation of $\epsilon$ and $\mu$, as the shape is determined by the combination $\epsilon\,\mu^{3/2}$, which is constrained by the CMB  anisotropies~\cite{James-Turner:2019ssu}. But, as is well known, the stochastic GW background which has a flat strength at a large frequency range can be probed at future GW detectors, which might reach $G\mu\sim10^{-18}$~\cite{Bian:2021vmi}. Therefore, if the asteroid-mass PBHs are from cosmic string collapse, then we might detect the associated GW signals, but have no hope to see the direct gamma-ray signals due to their low abundance.

\section{Summary and discussions}\label{sec:summary}

\begin{figure}
\centering
\includegraphics[scale=0.33]{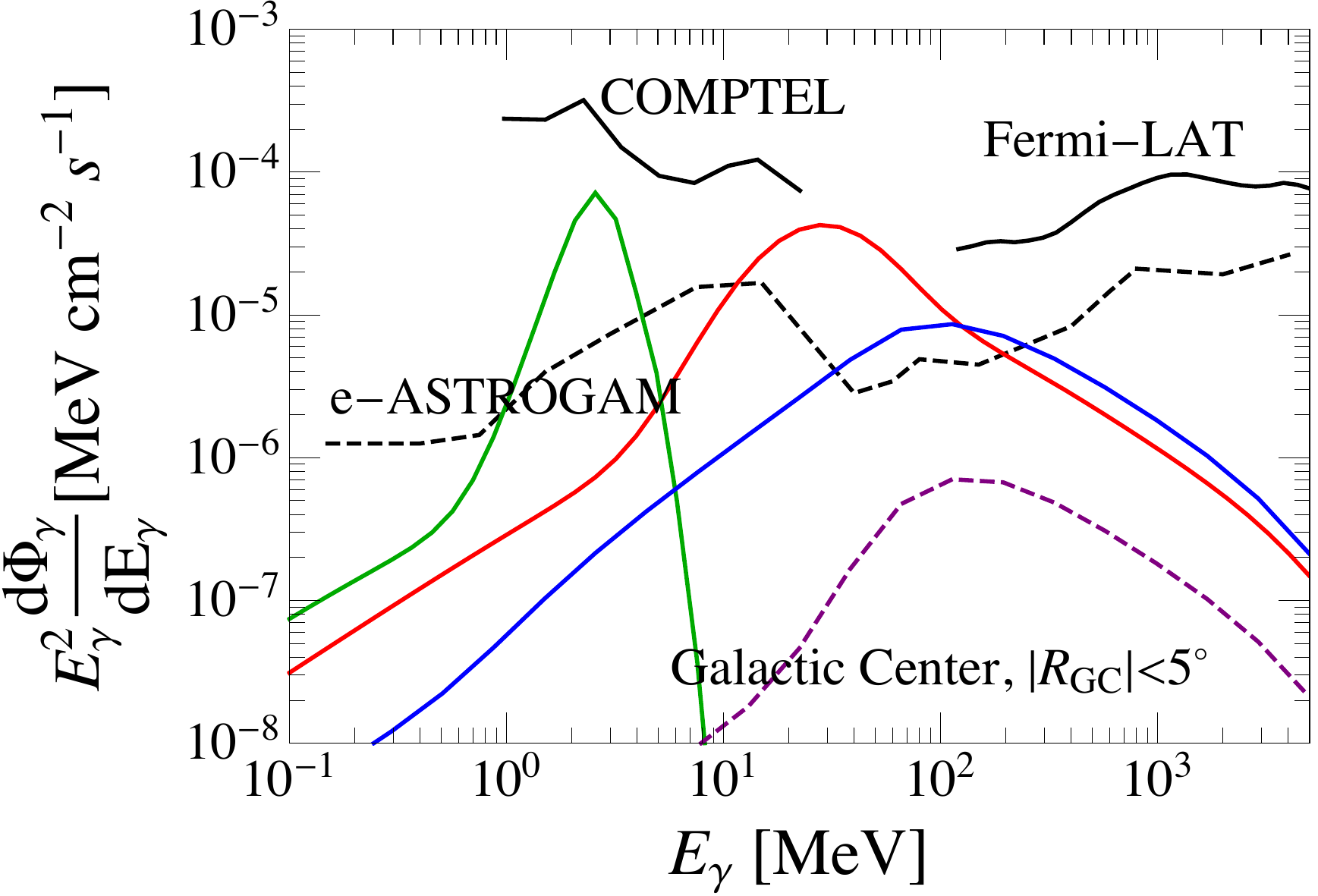}
\includegraphics[scale=0.32]{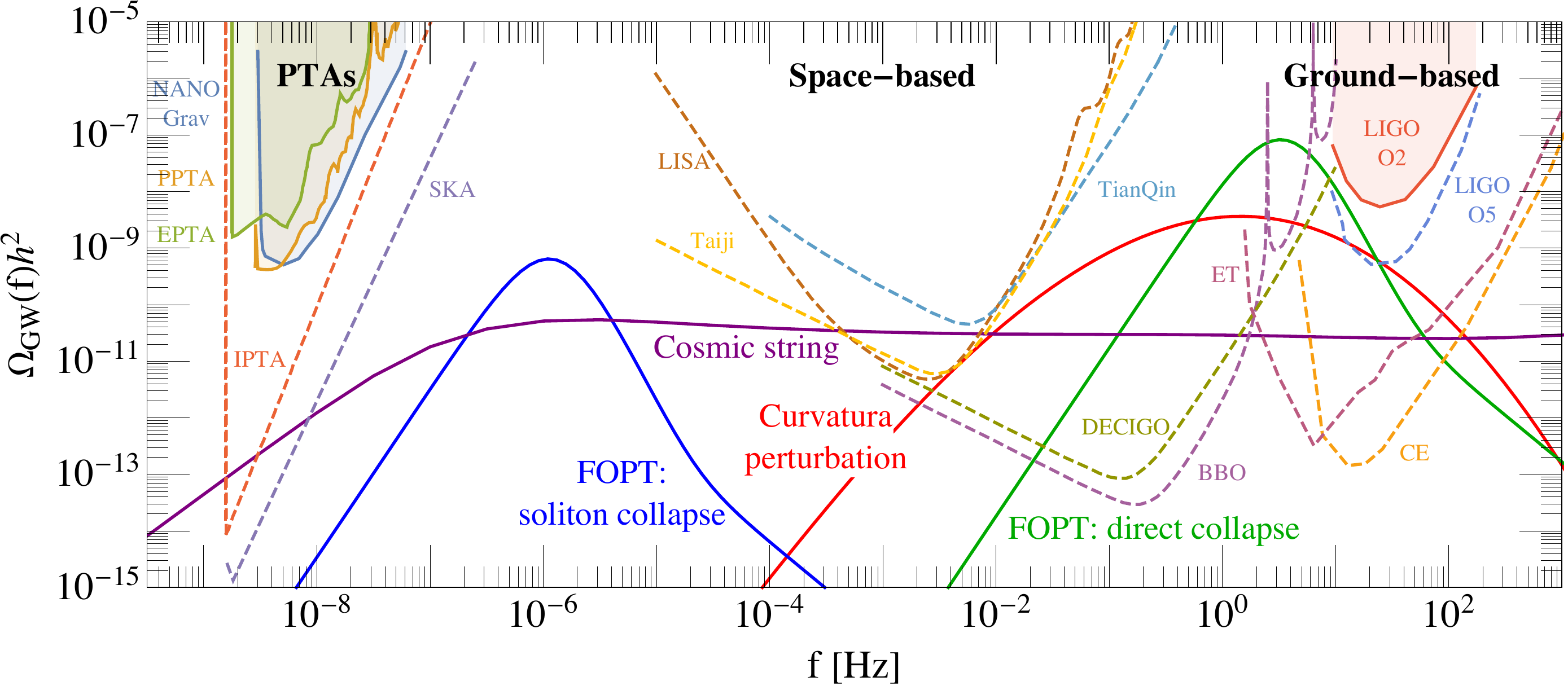}
\caption{The combined plot for gamma-rays ({\bf left}) and GWs ({\bf right}) of different PBH formation mechanisms. The red, green, blue and purple colors correspond to PBHs from curvature perturbation, direct collapse and soliton collapse in a FOPT, and cosmic strings, respectively. The gamma-ray signal strength from cosmic string-induced PBHs is multiplied by a factor of 100 for comparison.}
\label{fig:combined}
\end{figure}

We summarize the main features of the four PBH scenarios in Table~\ref{tab:features}. One can see that they have qualitative different features originating from different physics in formation, and hence could be distinguished experimentally by combining the gamma-ray and GW signals. The asteroid-mass PBHs from cosmic strings are already constrained to have a very low abundance and their Hawking radiation signals are much smaller than indirect detection backgrounds. For the other three mechanisms, the gamma-rays are reachable at the e-ASTROGAM~\cite{e-ASTROGAM:2016bph} and AMEGO-X~\cite{Fleischhack:2021mhc} detectors which are proposed to be launched in the late 2020s. Solely using the gamma-ray signals, one can distinguish the ``direct collapse during FOPT'' scenario from the other two scenarios, as the corresponding PBHs have very sharp mass functions that yield sharp gamma-ray spectra. Furthermore, as the three mechanisms have very different associated GW signals, they can be clearly classified with the help of the future GW detectors built in 2030s. The summary plots of gamma-ray and GW signals are given in Fig.~\ref{fig:combined}, where the first three scenarios use BP3 from \Eq{BPs_inflation}, \Eq{BPs_2105} and \Eq{BPs_kpx}, respectively, and for the cosmic string scenario we choose $w=10^{13}$ GeV.

\begin{table}\scriptsize\renewcommand\arraystretch{1.5}\centering
\begin{tabular}{c|c|c|c|c}\hline\hline
& Curvature perturbations & FOPT: direct collapse & FOPT: soliton collapse & Cosmic strings \\ \hline
Gamma-rays & Mild peak & Sharp peak & Mild peak & Not detectable \\ \hline
GW peak & $\mO(1)$ Hz & $\mO(10)$ Hz & $\mO(10^{-6})$ Hz & Flat \\ \hline\hline
\end{tabular} 
\caption{Signal features of different PBH formation mechanisms.}\label{tab:features}
\end{table}

PBH has been a hot topic in both astrophysics and particle physics, and many different PBH formation mechanisms have been proposed in the past several decades. Our research provides a first systematic comparison of four well-motivated and extensively studied mechanisms, pointing out their own characteristic features in the asteroid-mass PBH region. We have demonstrated that with the new instruments in the next decade, we can hopefully not only detect asteroid-mass PBHs, but also identify their origins. Our work can be extended further. There are other mechanisms of forming PBHs, such as bubble collisions~\cite{Hawking:1982ga,Kodama:1982sf,Moss:1994iq,Konoplich:1999qq,Kusenko:2020pcg,Jung:2021mku} or delayed vacuum decay~\cite{Liu:2021svg,Hashino:2021qoq,He:2022amv,Kawana:2022olo} from a FOPT, or the collapse of domain walls~\cite{Ferrer:2018uiu,Ge:2019ihf,Liu:2019lul}, see the review~\cite{Escriva:2022duf} for more mechanisms. Studying the features of those other mechanisms and the possibility of identifying them in experiments would be interesting. Besides, other channels from multi-messenger astronomy, such as the $e^\pm$ and neutrinos from PBH evaporation, can be also used to pin down the PBH formation mechanism. We leave those studies for future work.

\acknowledgments

We would like to thank Michael J. Baker, Anne M. Green, Joachim Kopp, and Tao Xu for the very useful and inspiring discussions. We are especially grateful to Tao Xu for communications on PBHs induced by curvature perturbations.

\appendix
\section{Hawking Radiation and Gamma-ray spectrum}\label{app:gamma-ray}

The huge gradient of the gravitational potential at the black hole horizon leads to particle production from the vacuum, which is known as the Hawking radiation process. The Hawking radiation could be described by the thermodynamics of a quasi-blackbody radiation spectrum, with an effective thermal temperature 
\be
T_{\rm pbh}= \frac{M_{\rm Pl}^2}{8 \pi M}\approx1.05~{\rm MeV}\times\left(\frac{10^{16}~{\rm g}}{M}\right),
\ee
which implies the dominant particles from the Hawking radiation of asteroid-mass PBHs are $\gamma$, $e^\pm$, $\mu^\pm$, and $\pi^{\pm,0}$. We assume only SM particles are produced by Hawking radiation. Recent studies found axion-like particles can also be produced and leave distinguishable features in gamma-ray spectra~\cite{Agashe:2022phd,Jho:2022wxd,Li:2022mcf}. The production of beyond SM particles does not change the conclusion of this study and we leave a dedicated spectrum analysis for the future.

The gravitational production rate of a particle species $i$ of mass $m_i$ in a logarithmic interval of energy $E_i$ is 
\beq
  \frac{\partial^2 N_{i,{\rm pri}}}{\partial E_i \partial t} = \frac{g_i}{2 \pi}\frac{\Gamma_i}{e^{E_i/T_{\rm pbh}}\pm 1},
\eeq
where the subscript ``pri'' indicates this is the primary production, and $g_i$ is the degree of freedom of the emitted particle. The $+/-$ sign is taken for fermion/boson production. Although the Hawking radiation spectrum follows almost a blackbody shape, the existence of re-absorption at the black hole horizon causes a small modification on the low energy part of the energy spectrum. The deviation from a pure blackbody can be parametrized in the so-called graybody factor $\Gamma_i$. The asymptotic approximation of graybody factor in the high energy limit $E_i\gg T_{\rm pbh}$ is $\Gamma_i=27 G^2 m^2_i E^2_i$, while numerical methods are required to determine the re-absorption rate in the low energy region. One should note that the re-absorption rate varies for particles with different spins. We refer to the package {\tt BLACKHAWK}~\cite{Arbey:2019mbc,Arbey:2021mbl} for the value of $\Gamma_i$ for different particles' spin and energy.

\begin{figure}
\centering
\includegraphics[scale=0.4]{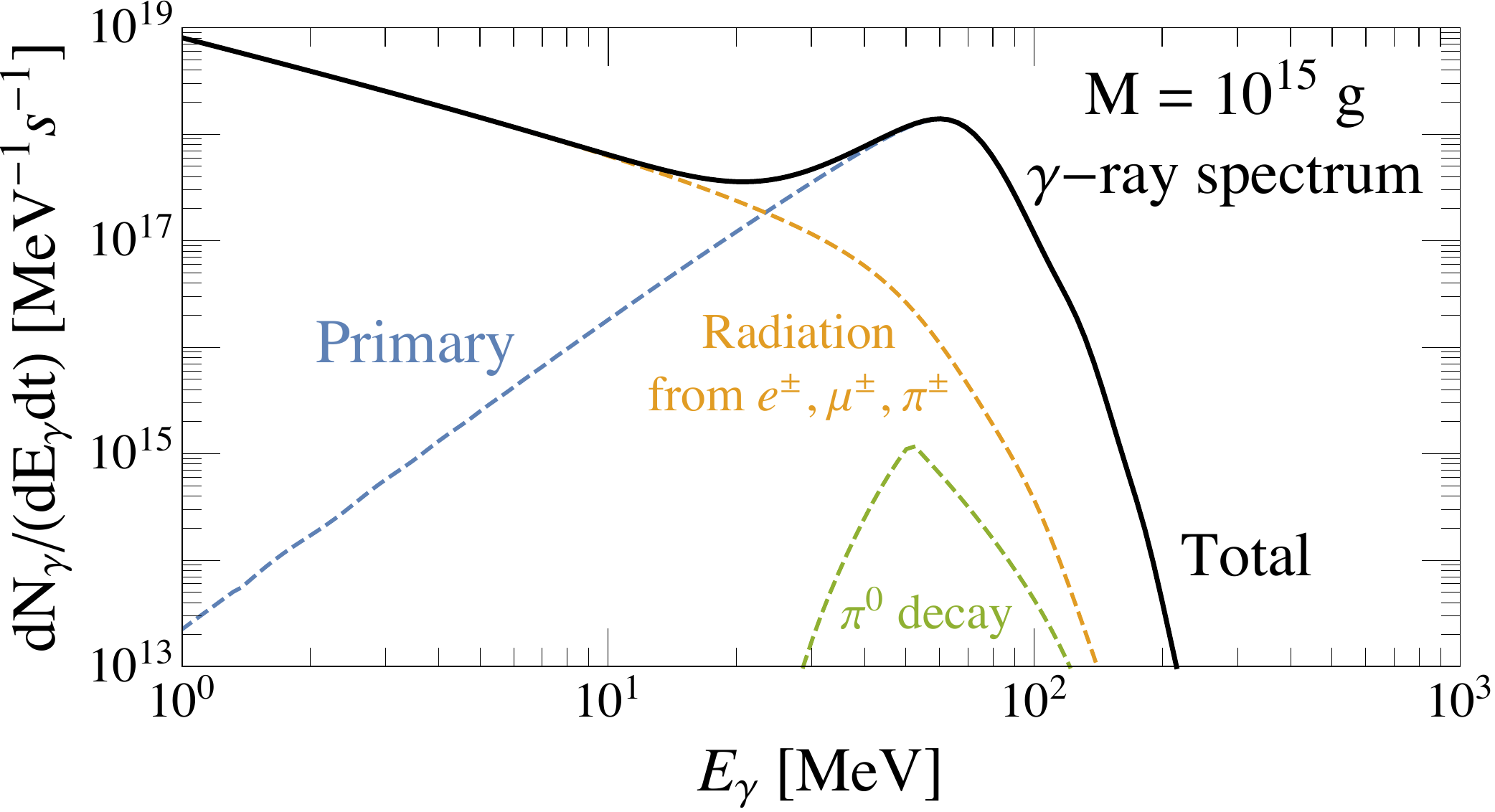}
\caption{The gamma-ray spectrum for a single PBH with $M=10^{15}$ g.}
\label{fig:hawking}
\end{figure}

The gamma-ray spectrum from a PBH is contributed by the primary photons and the secondary photons from final state radiation (FSR) or decay of other primary particles. For the FSR from $e^\pm$, $\mu^\pm$ and $\pi^{\pm}$, there is
\bea
    \frac{\d N_{i\to i\gamma}}{\d E_\gamma} &=& \frac{\alpha}{2\pi E_i}P_{i\rightarrow i\gamma}(x) \left [\log \left (\frac{1-x}{\mu_i^2} \right ) -1 \right ],\label{eq.FSR1}\\
    P_{i\rightarrow i\gamma}(x) &=& \begin{dcases}~ \frac{2(1-x)}{x}, & i=\pi^\pm; \\ ~\frac{1+(1-x)^2}{x}, & i=e^\pm,~\mu^\pm, \end{dcases}
\eea
where $x=E_\gamma/E_i$, $\mu_i=m_i/(2E_i)$. While for the decay from $\pi^0$, one obtains
\be
 \frac{\d N_{\pi^0\to\gamma\gamma}}{\d E_\gamma} = \frac{\Theta(E_\gamma-E_{\pi^0}^-) \Theta(E_{\pi^0}^+-E_\gamma)}{E_{\pi^0}^+-E_{\pi^0}^-},\quad
    E_{\pi^0}^\pm = \frac{1}{2} \left ( E_{\pi^0}\pm \sqrt{E_{\pi^0}^2 - m_{\pi^0}^2} \right ),
\ee
where $\Theta$ is the Heaviside function coming from the possible energy range of the final state photon. Summing up the above contributions, the total gamma-ray flux we have from a PBH is
\begin{multline}\label{eq:GammaRayTot}
\frac{\partial^2N_{\gamma}}{\partial E_\gamma\partial t}=\frac{\partial^2 N_{\gamma,{\rm pri}}}{\partial E_{\gamma}\partial t}\\
 + \sum_{i=e^\pm,\mu^\pm,\pi^\pm}\int\d E_i\frac{\partial^2N_{i,{\rm pri}}}{\partial E_i\partial t}\frac{\d N_{i\to i\gamma}}{\d E_\gamma} +2\int\d E_{\pi^0} \frac{\partial^2N_{\pi^0,{\rm pri}}}{\partial E_{\pi^0}\partial t}\frac{\d N_{\pi^0\to\gamma\gamma}}{\d E_\gamma}.
\end{multline}
The total spectrum is illustrated in Fig.~\ref{fig:hawking}. Primary photon production is the most important contribution to the total photon spectrum, due to the highest flux and the direct relation between the photon spectrum peak and the PBH temperature.

With the gamma-ray production rate from a single PBH in \Eq{eq:GammaRayTot}, we can calculate the total gamma-ray flux from the target source. In this study, we focus on using an observation towards the galactic center. The gamma-ray flux $\Phi_{\gamma}$ at the earth can be calculated as
\be\begin{split}
\frac{\d \Phi_\gamma}{\d E_{\gamma}}=&~\frac{\Delta \Omega}{4\pi} \times\left( \frac{1}{\Delta \Omega} \int_{\Delta \Omega} \d\Omega \int_{\rm los}\d l \, \rho_{\rm DM}\right) \int \frac{\d M}{M} \frac{\d f_{\rm pbh}}{\d M} \frac{\partial^2N_{\gamma}}{\partial E_{\gamma}\partial t}\\
\equiv&~\frac{\Delta \Omega}{4\pi} J_D \int \frac{\d M}{M} \frac{\d f_{\rm pbh}}{\d M} \frac{\partial^2N_{\gamma}}{\partial E_{\gamma}\partial t}.
\end{split}\ee
The $J_D$ is the D-factor of the galactic center for a ROI $\Delta\Omega$. We choose $\Delta\Omega$ to be $|R_{\rm GC}|<5^\circ$ so that $\Delta\Omega=2.39\times10^{-2}~{\rm sr}$. For the distribution of local DM abundance, we use an NFW distribution for the Milky Way with as~\cite{Navarro:1996gj} 
\beq
\rho_{\rm DM}(r)=\frac{\rho_s}{\frac{r}{r_s}\left(1+\frac{r}{r_s}\right)^{2}}.
\eeq
The parameters of the NFW profile is taken from Table 3 of Ref.~\cite{2019JCAP...10..037D} as $r_s=11~{\rm kpc}$ and $\rho_{{\rm DM},\odot}=0.376 ~{\rm GeV}/{\rm cm}^3$, which we use to derive $\rho_s=0.839~{\rm GeV}/{\rm cm}^3$. We use $r_{200}=193~{\rm kpc}$ to set the boundary of the galactic center halo for the line-of-sight (los) integral. In the end, we obtain the D-factor value $J_D=1.597 \times 10^{26}~{\rm MeV}\cdot {\rm cm}^{-2}\cdot{\rm sr}^{-1}$~\cite{Coogan:2020tuf}.

\bibliographystyle{JHEP-2-2.bst}
\bibliography{references}

\end{document}